\documentclass{aa}
\usepackage{graphicx}
\usepackage{amsmath}
\usepackage{threeparttable}
\usepackage{booktabs}
\usepackage{changepage}
\usepackage{ifthen}
\usepackage{xcolor}
\usepackage{txfonts}
\newboolean{showrevisions}
\setboolean{showrevisions}{true} % 切换 true/false
\newcommand{\revise}[1]{%
  \ifthenelse{\boolean{showrevisions}}{\textcolor{red}{#1}}{#1}%
}

\graphicspath{{./}{figures/}}

\begin{document}

\titlerunning{Radio emission from FBOTs}
\authorrunning{Liu et al.}

\title{Radio Emission from Fast Blue Optical Transients Powered by Trans-relativistic Shocks in Confined Circumstellar Material}

\author{
Liang-Duan Liu\inst{1,2,3}\corrauth{liuld@ccnu.edu.cn}
\and
Jia-Sen Zhang\inst{1}\email{jiasenzhang@mails.ccnu.edu.cn}
\and
Zhao-Sheng Zhang\inst{1}\email{zhangzs@mails.ccnu.edu.cn}
\and
Yun-Wei Yu\inst{1,2,3}\corrauth{yuyw@ccnu.edu.cn}
}

\institute{
Institute of Astrophysics, Central China Normal University, Wuhan 430079, China
\and
Laboratory for Compact Object Astrophysics and Astronomical Technology, Central China Normal University, Wuhan 430079, China
\and
Education Research and Application Center, National Astronomical Data Center, Wuhan 430079, China
}

\date{Received XXX; accepted XXX}
\abstract
{Fast blue optical transients (FBOTs) are luminous, rapidly evolving explosions whose radio emission provides a vital probe of shock dynamics and the density structure of the circumstellar material (CSM). Most existing radio studies rely on epoch-by-epoch spectral fits; however, this approach lacks dynamical consistency and fails to reproduce the steep post-peak fading observed in several well-sampled events.}
{We aim to develop a self-consistent physical framework for interpreting the radio emission of FBOTs and for constraining the immediate mass-loss history of their progenitors.}
{We present a forward-shock synchrotron model for ejecta interacting with a dense but radially confined CSM described by a broken power-law density profile. By incorporating synchrotron self-absorption and external free-free absorption, we model the multi-frequency radio light curves and broadband spectra of a representative sample of FBOTs.}
{We show that the diversity of FBOT radio morphologies can be naturally explained by shock interaction within a finite CSM shell: absorption regulates the early-time evolution, while the rapid post-peak decline marks the forward shock's exit from the dense inner region into a tenuous environment. The inferred shock velocities are trans-relativistic ($v \sim 0.1$--$0.5c$) and are associated with high mass-loading rates ($\dot{M} \sim 10^{-4}$--$10^{-3}\,M_{\odot}\,\mathrm{yr}^{-1}$), yet modest total CSM masses ($M_{\rm csm} \sim 10^{-4}$--$10^{-2}\,M_{\odot}$).}
{These properties point to brief episodes of enhanced mass loss occurring years to decades prior to explosion, rather than long-lived steady winds. Our framework provides a physically motivated interpretation of FBOT radio emission and a robust diagnostic for reconstructing the immediate mass-loss history of their progenitors.}


\maketitle
\section{Introduction}

Fast blue optical transients (FBOTs) are a class of luminous and rapidly evolving astronomical events. They exhibit high peak luminosities, blue colors, and short rise times \citep[e.g.,][]{Drout2014,Pursiainen2018}. These properties generally disfavor radioactive decay as the dominant power source. Most theoretical models instead invoke either a central engine, such as a millisecond magnetar \citep{Yu2015,Liu2022} or an accreting stellar-mass black hole \citep{Kashiyama2015}, or shock interaction with dense circumstellar material (CSM) \citep[e.g.,][]{Leung2020,Xiang2021,Pellegrino2022}. Tidal disruption events (TDEs) involving intermediate-mass black holes have also been proposed as a viable central-engine scenario, as they naturally account for the rapid timescales and high X-ray luminosities observed in some FBOTs \citep[e.g.,][]{Perley2019,Kremer2021}. Although these models can reproduce the observed thermal optical emission, they predict distinct outflow dynamics and environmental conditions. Diagnostics beyond the optical band are therefore required to discriminate among them.

Radio observations provide a powerful probe of ejecta dynamics and the CSM, offering constraints that are inaccessible to thermal emission alone. In high-energy transients, radio emission typically arises from synchrotron radiation produced by shock-accelerated electrons in the local magnetic field. Its temporal evolution is governed by the shock velocity and absorption processes, including synchrotron self-absorption (SSA) and free–free absorption (FFA). As a result, radio emission directly traces the mass loading of the environment and constrains the progenitor’s mass-loss history \citep[e.g.,][]{Chevalier1998,Weiler2002}. Despite this diagnostic power, radio detections of FBOTs remain rare. In the events with radio detections, the light curves often show complex behavior, including luminous peaks and unexpectedly steep decay phases \citep[e.g.,][]{Ho2019,Coppejans2020}.

This limitation is likely driven by instrumental sensitivity and follow-up selection effects, as intensive radio monitoring is typically triggered only for the most luminous events or those with prominent high-energy emission \citep{Margutti2019,Bright2022,Yao2022_AT2020mrf}. Recent population synthesis studies further show that current radio facilities, such as the VLA, are sensitive to only a small fraction of the optical FBOT population ($\sim9\%$), naturally accounting for the limited size of the existing radio sample \citep{Liu2023}.

Within the observed sample, FBOT radio spectral luminosities generally lie between those of ordinary core-collapse supernovae and canonical GRB afterglows. Most previous studies have adopted an epoch-by-epoch approach, fitting the radio spectral energy distribution (SED) at each epoch with an empirical broken power law to represent a self-absorbed synchrotron spectrum. The shock radius and magnetic field are then inferred from the peak flux and peak frequency at each epoch \citep[e.g.,][]{Chevalier1998}. Although this method provides convenient instantaneous estimates, it treats each epoch independently. As a result, the inferred parameter evolution may not follow dynamically consistent shock scalings or robustly constrain the continuous CSM density profile. Moreover, the steep post-peak fading observed in several FBOTs is difficult to reproduce with standard steady-wind models, instead favoring a dense CSM that is radially confined.

In this work, we overcome these limitations by developing a self-consistent physical framework for modeling multi-frequency radio light curves of FBOTs.
Unlike snapshot SED fitting, our approach adopts a single, physically coupled parameter set to describe the shock dynamics, magnetic-field evolution, and CSM structure, thereby reproducing the temporal and spectral evolution in a unified manner.
By incorporating a broken power-law CSM density profile into the forward-shock dynamics, the model naturally captures both the optically thick phase and the subsequent steep optically thin decline expected from radially confined CSM.
We apply this framework to multi-frequency radio light curves of ZTF18abvkwla, AT2020xnd, and AT2024wpp, as well as to broadband radio SEDs of CSS161010, AT2020mrf, and AT2023fhn.
The shock velocity, mass-loading rate, and CSM density structure are constrained through a Bayesian inference analysis.

This paper is organized as follows. Section~\ref{sec:radio_obs} summarizes the radio observational properties of FBOTs. Section~\ref{sec:framework} presents the physical model. Section~\ref{sec:result} explores the resulting light-curve behavior and parameter dependence. Section~\ref{sec:Application} applies the model to individual events and presents the inferred physical properties. Section~\ref{sec:discussion} discusses the implications, predictions, and observational strategies, and summarizes our conclusions.

\section{Radio observational properties of FBOTs}
\label{sec:radio_obs}

\begin{table*}[t]
\centering
\caption{Observed Radio Peak Properties of FBOTs}
\label{tab:radio}
\footnotesize
\renewcommand{\arraystretch}{1.2} 
\setlength{\tabcolsep}{6pt}       

\begin{tabular}{l c c c c c l}
\toprule
\toprule
Transient & $z$ & $F_{\nu,\mathrm{pk}}$ & $L_{\nu,\mathrm{pk}}$ & $t_{\mathrm{pk}}$ & $\nu_{\mathrm{pk}}$ & References \\
 & & (mJy) & ($10^{29}$ erg s$^{-1}$ Hz$^{-1}$) & (days) & (GHz) & \\
\midrule

AT2018cow
& 0.0141 & 93.97 & 2.78 & 22.0  & 102.5 & \cite{Ho2019} \\
&        & 9.80  & 0.46 & 83.5  & 11.0  & \cite{Margutti2019} \\
&        & 2.88  & 0.14 & 138.8 & 1.2   & \cite{Nayana2021_AT2018cow} \\
\addlinespace 

ZTF18abvkwla
& 0.2700 & 0.36  & 6.90 & 81.0  & 10.0  & \cite{Ho2020} \\
\addlinespace

CSS161010
& 0.0336 & 7.06  & 1.90 & 99.0  & 5.4   & \cite{Coppejans2020} \\
\addlinespace

AT2020xnd
& 0.2433 & 0.49  & 7.49 & 48.2  & 34.0  & \cite{Bright2022} \\
&        & 1.08  & 16.44& 31.8  & 79.0  & \cite{Ho2022_AT2020xnd} \\
\addlinespace

AT2020mrf
& 0.1353 & 0.35  & 1.59 & 259.5 & 7.7   & \cite{Yao2022_AT2020mrf} \\
\addlinespace

AT2022tsd
& 0.2564 & 0.57  & 9.78 & 16.9  & 207.3 & \cite{Ho2023_AT2022tsd_Nature} \\
\addlinespace

AT2023fhn
& 0.2380 & 0.22  & 3.23 & 137.1 & 6.0   & \cite{Chrimes2024_AT2023fhn} \\
\addlinespace

AT2024wpp
& 0.0868 & 1.28  & 2.36 & 35.8  & 97.5  & \cite{Nayana2025_AT2024wpp} \\
&        & 1.50  & 2.77 & 44.9  & 33.0  & \cite{Perley2026_AT2024wpp} \\

\bottomrule
\end{tabular}

\begin{flushleft}
\footnotesize
\textbf{Note.} Columns list the redshift ($z$), peak flux density ($F_{\nu,\mathrm{pk}}$),
peak spectral luminosity ($L_{\nu,\mathrm{pk}}$), observer-frame time of the radio peak
($t_{\mathrm{pk}}$), and the corresponding observing frequency
($\nu_{\mathrm{pk}}$).

\end{flushleft}
\end{table*}

\subsection{Luminosity and spectral morphology}
Radio follow-up observations show that the most radio-bright FBOTs are luminous
and rapidly evolving transients at cm–mm wavelengths. Table~\ref{tab:radio}
summarizes the observed radio peak properties of FBOTs with published
multi-frequency detections, including the peak flux density, spectral
luminosity, peak epoch, and observing frequency. The sample spans a wide range
of peak timescales and reaches high radio luminosities. At GHz frequencies,
FBOTs typically attain peak spectral luminosities of
$L_{\nu,\mathrm{pk}}\sim10^{28}$–$10^{30}\ {\rm erg\ s^{-1}\ Hz^{-1}}$, placing
them well above ordinary core-collapse supernovae
\citep{Ho2019,Coppejans2020}. While generally fainter than on-axis long-GRB
afterglows, FBOTs overlap with the low-luminosity end of the GRB radio
population \citep{Bright2022}.

Multi-frequency radio SEDs are often well
described by a peaked broken power law, consistent with synchrotron emission
affected by SSA. A representative example is
CSS~161010, in which the SED peak shifts to lower frequencies with time; at
$t\sim99$~days, the optically thick and thin spectral indices are
$\alpha=2.00\pm0.08$ and $\alpha=-1.31\pm0.03$
($F_{\nu}\propto\nu^{\alpha}$), respectively \citep{Coppejans2020}. Similarly,
AT~2020mrf shows a broken power-law SED with $\nu_{\rm pk}\approx7.4$~GHz and an
optically thin slope of $\alpha\approx-1.0$ at late times
\citep{Yao2022_AT2020mrf}.

While the standard homogeneous SSA model, which predicts an optically thick
slope of $\alpha=2.5$, provides a useful baseline \citep{Chevalier1998},
deviations are frequently observed. AT~2024wpp exhibits unusually flat
optically thick spectra, with $\alpha\approx0.7$–1.5 \citep{Nayana2025_AT2024wpp},
which may indicate geometric effects or source inhomogeneity. In contrast,
constraints on extrinsic absorption processes, such as FFA, are often strong. In both AT~2018cow and AT~2020xnd, the lack of a sharp
low-frequency spectral cutoff implies a low density for any extended, ionized
absorbing medium \citep{Ho2019,Ho2022_AT2020xnd,Bright2022}.

Millimeter and submillimeter observations reveal additional complexity in the
early-time radio SEDs. AT~2018cow showed a rapid rise at early epochs, peaking
within $\sim3$ weeks, with 90–100~GHz flux densities reaching
$\sim94$~mJy \citep{Ho2019}. More recently, AT~2024wpp displayed a highly
non-standard evolution: the relative ordering of the 97.5 and 203~GHz flux
densities reversed between $t\approx19$ and 36~days, tracing the passage of the
spectral peak through the millimeter band. At later times ($t\gtrsim100$~days),
AT~2024wpp developed a spectral inversion, which has been described as
unprecedented and may signal the emergence of an additional emission
component \citep{Perley2026_AT2024wpp}.

\subsection{Temporal evolution}

FBOT radio emission is typically detected at early epochs and rises rapidly to
peak brightness. The time to peak is generally short, ranging from a few days
to a few tens of days depending on observing frequency, with higher frequencies
peaking earlier than lower ones. This behavior contrasts with the slower,
months-long evolution commonly seen in ordinary radio supernovae.

A prominent observational feature in many well-sampled FBOTs is an
exceptionally steep post-peak decline. In CSS~161010, following a
frequency-dependent peak (e.g., at $t\simeq99$~days at 5.36~GHz), the radio
luminosity declines as $L_{\nu}\propto t^{-5.1\pm0.3}$ \citep{Coppejans2020}. In
ZTF18abvkwla, the 10~GHz flux density decreases by roughly an order of magnitude
between 81 and 396~days, consistent with a decay of
$F_{\nu}\propto t^{-3.2\pm1.4}$ \citep{Ho2020}. Similarly, AT~2020xnd exhibits
rapid fading after an early millimeter peak, with the post-peak evolution
empirically described by $F_{\nu}\propto t^{-4}$
\citep{Ho2022_AT2020xnd}. These decay indices are substantially steeper than the
$t^{-1}$–$t^{-1.5}$ declines typically observed in ordinary radio supernovae.

Not all FBOTs follow this trend of rapid fading. AT~2024wpp shows a more
moderate decline, with a decay slope of $F_{\nu}\propto t^{-2.05}$, while
AT~2022tsd exhibits a relatively flat radio light curve without a clear
monotonic trend over the observed time span. For AT~2023fhn, the available
radio data primarily sample the rising phase around $t\sim100$~days, and the
source has not yet entered a clear post-peak decline
\citep{Chrimes2024_AT2023fhn}. These cases illustrate that FBOT radio light
curves can display a range of post-peak behaviors, particularly when temporal
sampling is sparse or restricted to specific evolutionary phases.

Nevertheless, very steep decay indices are prevalent among the best-sampled
events, including CSS~161010, ZTF18abvkwla, and AT~2020xnd, indicating that
rapid post-peak fading is a common characteristic of radio-bright FBOTs.
Observationally, this behavior implies a short-lived phase of strong radio
emission and is consistent with scenarios in which efficient shock interaction
terminates once the shock exits a radially confined CSM environment.

\section{Physical framework} \label{sec:framework}
We present a self-consistent framework in which FBOT radio emission arises
from synchrotron radiation produced by a forward shock driven by the
interaction between trans-relativistic ejecta, with velocities of order a few
tenths of the speed of light, and the surrounding circumstellar medium.
In this scenario, the fast outflow drives a shock into the CSM, and
a fraction of the post-shock energy is converted into non-thermal electrons
and magnetic fields, generating broadband synchrotron emission. The observed
radio light curves are shaped by early-time absorption, primarily synchrotron
self-absorption within the emitting region and, when present, free–free
absorption in the external ionized CSM, together with the intrinsic decline of
the interaction power as the shock expands. To account for indications that
the dense environment is radially confined, we allow the CSM density profile
to transition at a characteristic radius. After the shock reaches this
transition, the effective target density decreases, and the radio emission
can fade more steeply, reproducing the sharp post-peak declines observed in
some FBOTs. With a small set of physical parameters, the model predicts the
full multi-frequency evolution $F_{\nu}(t)$, enabling a unified
interpretation of radio morphologies across epochs rather than relying on
isolated, epoch-by-epoch estimates.

\subsection{Shock dynamics from ejecta--CSM interaction}
\label{subsec:dynamic}

The interaction between the freely expanding ejecta and the CSM begins when the fastest ejecta reaches the inner edge of the CSM, $R_{\rm csm,in}$. At small radii, the interaction region is generally optically thick to Thomson scattering, and the forward shock is expected to be radiation mediated. In this regime, the strong coupling between electrons and radiation promotes rapid thermalization and suppresses the formation of a long-lived non-thermal power-law electron population. As the shock propagates outward, the optical depth ahead of the shock decreases. After shock breakout at $R_{\rm bo}$, the shock becomes collisionless \citep[e.g.,][]{Katz2010,Murase2011}, enabling efficient diffusive shock acceleration and the sustained production of non-thermal electrons. We therefore identify the onset of the collisionless-shock phase with the beginning of the radio-emitting stage.

% We define the effective initial radius as
% \begin{equation}
% R_{\rm in}\equiv \max(R_{\rm csm,in},R_{\rm bo}),
% \end{equation}
% and the corresponding start time as
% \begin{equation}
% t_{\rm in}\equiv \frac{R_{\rm in}}{v_{\rm in}}
% \simeq 0.27~{\rm days}~
% R_{{\rm in},3}v_{{\rm in},-1}^{-1},
% \label{eq:tin_def}
% \end{equation}
% where $v_{\rm in}$ is the forward-shock velocity at $R_{\rm in}$,
% $R_{{\rm in},3}\equiv R_{\rm in}/10^{3}R_\odot$, and
% $v_{{\rm in},-1}\equiv v_{\rm in}/(0.1c)$. All radio-emission and absorption calculations in this work are restricted to $t\ge t_{\rm in}$.

\paragraph{CSM density structure.}
We adopt a spherically symmetric CSM described by a broken power-law density profile,
\begin{equation}
\rho_{\rm csm}(r)=
\begin{cases}
\rho_{\rm in}
\left(\dfrac{r}{R_{\rm in}}\right)^{-s_1},
& R_{\rm in}\le r < R_{\rm br},\\[6pt]
\rho_{\rm br}
\left(\dfrac{r}{R_{\rm br}}\right)^{-s_2},
& r \ge R_{\rm br},
\end{cases}
\label{eq:csm_broken_powerlaw}
\end{equation}
where $\rho_{\rm in}\equiv \rho_{\rm csm}(R_{\rm in})$ is the density normalization at the effective inner radius. Continuity at the break radius requires
\begin{equation}
\rho_{\rm br}
=
\rho_{\rm in}
\left(\frac{R_{\rm br}}{R_{\rm in}}\right)^{-s_1}.
\end{equation}
The indices $s_1$ and $s_2$ describe the inner and outer CSM density slopes, respectively. The dense inner CSM controls the early absorption and interaction power, whereas the outer profile determines how rapidly the environment becomes tenuous after the shock crosses $R_{\rm br}$. A steep outer slope, $s_2\gg s_1$, effectively mimics a truncated CSM and can naturally produce a rapid post-peak decline in the radio light curve.

For the special case $s_1=2$, the inner CSM corresponds to a steady wind. The density normalization is then
\begin{equation}
\rho_{\rm in}
=
\frac{\dot{M}}{4\pi v_{\rm w}R_{\rm in}^{2}},
\label{eq:rho_in_wind}
\end{equation}
where $\dot{M}$ is the progenitor mass-loss rate and $v_{\rm w}$ is the wind velocity. The break radius can be connected to the pre-explosion mass-loss history through the look-back time,
\begin{equation}
t_{\rm pre,br}
\simeq
\frac{R_{\rm br}}{v_{\rm w}}
\simeq
22.1~{\rm yr}~
R_{{\rm br},6}v_{{\rm w},3}^{-1},
\label{eq:tpre_br}
\end{equation}
where $R_{{\rm br},6}\equiv R_{\rm br}/(10^{6}R_\odot)$ and
$v_{{\rm w},3}\equiv v_{\rm w}/(10^{3}~{\rm km~s^{-1}})$.

\paragraph{Shock evolution.}
We describe the dynamical evolution of the interaction region using the thin-shell approximation. In this framework, the shocked CSM and shocked outer ejecta are treated as a geometrically thin shell with radius $R_{\rm sh}$, velocity $v_{\rm sh}={\rm d}R_{\rm sh}/{\rm d}t$, and swept-up mass $M_{\rm sh}$. This approximation captures the leading-order momentum exchange between the freely expanding ejecta and the CSM without resolving the internal structure between the forward shock, contact discontinuity, and reverse shock.

For homologously expanding ejecta, the unshocked ejecta velocity at the shell position is
\begin{equation}
v_{\rm ej}(R_{\rm sh},t)=\frac{R_{\rm sh}}{t}.
\end{equation}
The shell mass grows by sweeping up both the CSM and the faster ejecta that overtakes the shell. Assuming an initially stationary CSM, the mass growth rate is
\begin{equation}
\frac{{\rm d}M_{\rm sh}}{{\rm d}t}
=
4\pi R_{\rm sh}^{2}
\left[
\rho_{\rm csm}(R_{\rm sh})v_{\rm sh}
+
\rho_{\rm ej}(R_{\rm sh},t)
\left(
\frac{R_{\rm sh}}{t}-v_{\rm sh}
\right)
\right],
\label{eq:dMsh_dt}
\end{equation}
where the ejecta term contributes only when $R_{\rm sh}/t>v_{\rm sh}$.

The shell momentum evolves as
\begin{equation}
\begin{aligned}
\frac{{\rm d}}{{\rm d}t}
\left(M_{\rm sh}v_{\rm sh}\right)
&=
4\pi R_{\rm sh}^{2}
\bigg[
\rho_{\rm ej}(R_{\rm sh},t)
\left(
\frac{R_{\rm sh}}{t}-v_{\rm sh}
\right)^{2}
\\
&\quad
-
\rho_{\rm csm}(R_{\rm sh})v_{\rm sh}^{2}
\bigg].
\end{aligned}
\label{eq:thin_shell_momentum}
\end{equation}
The first term on the right-hand side is the ram pressure supplied by the faster ejecta, while the second term is the opposing ram pressure from the swept-up CSM.

For the freely expanding outer ejecta, we adopt a homologous power-law density profile,
\begin{equation}
\rho_{\rm ej}(r,t)\propto t^{-3}v^{-n},
\qquad
v=\frac{r}{t},
\end{equation}
where $n$ is the outer ejecta density index. In Equations~(\ref{eq:dMsh_dt}) and~(\ref{eq:thin_shell_momentum}), this profile is evaluated at $r=R_{\rm sh}(t)$. The CSM density is given by Equation~(\ref{eq:csm_broken_powerlaw}), allowing the same set of equations to be integrated continuously across $R_{\rm br}$. The transition time $t_{\rm br}$ is defined by $R_{\rm sh}(t_{\rm br})=R_{\rm br}$.

In the numerical thin-shell treatment, $t_{\rm br}$ is obtained directly by integrating Equations~(\ref{eq:dMsh_dt}) and~(\ref{eq:thin_shell_momentum}) together with $v_{\rm sh}={\rm d}R_{\rm sh}/{\rm d}t$. Once the shock crosses $R_{\rm br}$, the rapid decline in the external density reduces both the swept-up mass and the kinetic-energy dissipation rate, leading naturally to a steepening of the radio light curve.

In the light-curve calculations below, we use this numerical thin-shell solution as the fiducial dynamical model. The analytic self-similar solution is introduced only as a limiting case for single power-law ejecta and CSM density profiles. It is used to provide physical intuition and to define characteristic shock-evolution timescales, rather than as the primary dynamical prescription used in the numerical modeling.

For reference, in the limiting case where both the outer ejecta and the CSM are described by single power-law density profiles, the thin-shell solution approaches the standard self-similar form. Normalizing the solution at the effective initial radius $R_{\rm in}$, we write
\begin{equation}
R_{\rm sh}(t)
=
R_{\rm in}
\left(
\frac{t}{t_{\rm in}}
\right)^{m},
\qquad
m=\frac{n-3}{n-s},
\qquad
t_{\rm in}\le t<t_{\rm br},
\label{eq:Rsh_powerlaw}
\end{equation}
where $s$ is the CSM density index, and $t_{\rm in}\equiv R_{\rm in}/v_{\rm in}$ is the time at which the shock reaches $R_{\rm in}$ \citep{Chevalier1982,Moriya2013}. This approximation gives
\begin{equation}
t_{\rm br}
=
t_{\rm in}
\left(
\frac{R_{\rm br}}{R_{\rm in}}
\right)^{1/m}
\simeq
720~{\rm days}~
R_{{\rm in},3}^{(m-1)/m}
R_{{\rm br},6}^{1/m}
v_{{\rm in},-1}^{-1}.
\label{eq:tbr_powerlaw}
\end{equation}
After the shock enters the lower-density outer region, its additional deceleration becomes weak. We therefore approximate the late-time evolution as nearly coasting,
\begin{equation}
R_{\rm sh}(t)
=
R_{\rm br}
\left(
\frac{t}{t_{\rm br}}
\right),
\qquad
t\ge t_{\rm br}.
\label{eq:Rsh_coast}
\end{equation}
This piecewise expression provides a compact analytic approximation to the numerical thin-shell dynamics: a mildly decelerating phase in the dense inner CSM, followed by nearly free expansion after the shock crosses the density break.

\subsection{Post-shock energy density and microphysics}
\label{subsec:microphysics}

We model the radio emission as synchrotron radiation from non-thermal electrons accelerated at the forward shock after breakout. In this work, we focus on the forward-shock contribution. For a steep outer ejecta profile, the reverse-shock contribution to the observed non-thermal emission is expected to be subdominant \citep{Chevalier2006}.

For a strong, sub-relativistic shock propagating into an upstream CSM density $\rho_{\rm csm}$, the post-shock internal energy density is approximated by the Rankine--Hugoniot jump conditions as
\begin{equation}
e_{\rm int}
\simeq
\frac{9}{8}
\rho_{\rm csm}(R_{\rm sh})
v_{\rm sh}^{2},
\label{eq:eint_98}
\end{equation}
where the factor $9/8$ corresponds to the strong-shock limit for an ideal gas with adiabatic index $\gamma_{\rm ad}=5/3$.

We parameterize the post-shock microphysics by assigning fixed fractions of the internal energy density to non-thermal electrons and magnetic fields,
\begin{equation}
e_{\rm e}=\epsilon_{\rm e}e_{\rm int},
\qquad
e_{\rm B}=\epsilon_{\rm B}e_{\rm int},
\end{equation}
where $\epsilon_{\rm e}$ and $\epsilon_{\rm B}$ are treated as phenomenological microphysical parameters. The magnetic-field strength in the emitting region is then
\begin{equation}
B=\left(8\pi e_{\rm B}\right)^{1/2}.
\end{equation}
For a wind-like inner CSM profile, $s_1=2$, this gives
\begin{equation}
B
\simeq
51.3~{\rm G}~
\epsilon_{B,-2}^{1/2}
\dot{M}_{-4}^{1/2}
v_{{\rm w},3}^{-1/2}
R_{{\rm in},3}^{-1}
v_{{\rm in},-1}
\left(
\frac{t}{t_{\rm in}}
\right)^{-1},
\label{eq:B_field_wind}
\end{equation}
where
$\epsilon_{B,-2}\equiv \epsilon_{\rm B}/10^{-2}$ and
$\dot{M}_{-4}\equiv \dot{M}/(10^{-4}~M_\odot~{\rm yr^{-1}})$.

Because particle acceleration and magnetic-field amplification in trans-relativistic SN shocks remain uncertain, $\epsilon_{\rm e}$, $\epsilon_{\rm B}$, and the accelerated-electron fraction $\xi_{\rm e}$ are treated as effective parameters.

We assume that a fraction $\xi_{\rm e}$ of the post-shock electrons is accelerated into a power-law distribution in Lorentz factor,
\begin{equation}
\frac{{\rm d}n_{\rm e}}{{\rm d}\gamma}
=
K_{\rm e}\gamma^{-p},
\qquad
\gamma\ge \gamma_{\rm m},
\label{eq:electron_pl}
\end{equation}
where $p$ is the electron spectral index, $\gamma_{\rm m}$ is the minimum Lorentz factor, and $K_{\rm e}$ is the normalization per unit volume.

For a strong shock with $\gamma_{\rm ad}=5/3$, the downstream number density is $n_2\simeq 4n_1$, where
\begin{equation}
n_1=\frac{\rho_{\rm csm}}{\mu m_{\rm p}}
\end{equation}
is the upstream number density, $\mu$ is the mean molecular weight, and $m_{\rm p}$ is the proton mass. Number conservation for the accelerated electrons gives
\begin{equation}
\int_{\gamma_{\rm m}}^\infty
\frac{{\rm d}n_{\rm e}}{{\rm d}\gamma}
\,{\rm d}\gamma
=
\xi_{\rm e}n_2,
\end{equation}
and hence
\begin{equation}
K_{\rm e}
=
(p-1)\xi_{\rm e}n_2\gamma_{\rm m}^{p-1}.
\label{eq:Ke_norm}
\end{equation}

The minimum Lorentz factor is determined by the electron energy budget,
\begin{equation}
\int_{\gamma_{\rm m}}^\infty
(\gamma-1)m_{\rm e}c^2
\frac{{\rm d}n_{\rm e}}{{\rm d}\gamma}
\,{\rm d}\gamma
=
e_{\rm e}.
\end{equation}
For $p>2$, this yields
\begin{equation}
\gamma_{\rm m}-1
\simeq
\frac{p-2}{p-1}
\frac{e_{\rm e}}{\xi_{\rm e}n_2m_{\rm e}c^2}.
\label{eq:gamma_m_general}
\end{equation}
This expression gives the standard scaling
$\gamma_{\rm m}-1\propto \epsilon_{\rm e}(m_{\rm p}/m_{\rm e})(v_{\rm sh}^{2}/c^{2})$ up to factors of order unity.

When radiative cooling is relevant, we define the synchrotron cooling Lorentz factor by equating the synchrotron cooling time to the dynamical time,
\begin{equation}
\gamma_{\rm c}
=
\frac{6\pi m_{\rm e}c}
{\sigma_{\rm T}B^{2}t},
\label{eq:gammac}
\end{equation}
where $\sigma_{\rm T}$ is the Thomson cross section. Electrons with $\gamma>\gamma_{\rm c}$ cool efficiently within a dynamical time, causing the distribution to steepen above $\gamma_{\rm c}$. Electrons with $\gamma<\gamma_{\rm c}$ are weakly affected by synchrotron losses and retain the injected slope.

For the parameter space relevant to radio SNe and FBOTs, the magnetic field typically decreases rapidly with time. The system is therefore usually in the slow-cooling regime, $\gamma_{\rm c}\gg \gamma_{\rm m}$. In this regime, the electrons responsible for GHz emission generally satisfy $\gamma\ll\gamma_{\rm c}$, so synchrotron cooling can be neglected in our radio light-curve calculations.

\subsection{Synchrotron emission and absorption processes}
\label{subsec:synch_abs}

For $t \ge t_{\mathrm{in}}$, we model the radio emission as synchrotron
radiation from non-thermal electrons in the shocked CSM immediately behind the
forward shock. The post-shock magnetic-field strength $B(t)$ and the electron distribution are specified in Section~\ref{subsec:microphysics}.  In this subsection we summarize
how the volume emissivity and absorption coefficients are converted into the
emergent specific intensity, and then into the observable luminosity,
including SSA and external FFA.

\subsubsection{Synchrotron intensity, absorption, and luminosity}
The emergent synchrotron specific intensity from the emitting shell follows from the formal solution of the radiative transfer equation,
\begin{equation}
I_{\nu,{\rm syn}} = S_{\nu,{\rm syn}}\left(1-e^{-\tau_{\nu,{\rm syn}}}\right),
\label{eq:Inu_syn}
\end{equation}
where
\begin{equation}
S_{\nu,{\rm syn}} \equiv \frac{j_{\nu,{\rm syn}}}{\alpha_{\nu,{\rm syn}}}
\label{eq:Snu_syn}
\end{equation}
is the synchrotron source function. Here $j_{\nu,{\rm syn}}$ and $\alpha_{\nu,{\rm syn}}$ are the synchrotron volume emissivity and absorption coefficient, respectively. In this work, we compute $j_{\nu,{\rm syn}}$ and $\alpha_{\nu,{\rm syn}}$ using the standard synchrotron formalism for a power-law electron distribution following \citealt{Rybicki1979}. The SSA optical depth across the emitting region is approximated as
\begin{equation}
\tau_{\nu,{\rm syn}} \simeq \alpha_{\nu,{\rm syn}} \, \Delta R,
\label{eq:tau_syn}
\end{equation}
where $\Delta R=\eta R_{\rm sh}$ is the effective thickness of the shocked emitting layer. At early times the forward shock is expected to be strong, with a compression ratio of $\simeq 4$  \citep[e.g.,][]{Sturner1997}. For a geometrically thin shocked shell, this motivates adopting a fiducial fractional width $\eta\simeq 0.25$.

If the unshocked CSM ahead of the shock acts as an external absorbing screen, the observed intensity is further attenuated by free--free absorption,
\begin{equation}
I_\nu = I_{\nu,{\rm syn}}\,e^{-\tau_{\nu,{\rm ff}}}
= S_{\nu,{\rm syn}}\left(1-e^{-\tau_{\nu,{\rm syn}}}\right)e^{-\tau_{\nu,{\rm ff}}},
\label{eq:Inu_total}
\end{equation}
where $\tau_{\nu,{\rm ff}}$ is the free--free optical depth integrated through the unshocked ionized CSM along the line of sight.
The free--free absorption coefficient can be written as
\begin{equation}
\alpha_{\nu,{\rm ff}} \simeq 0.018\,T_{\rm{e}}^{-3/2}\,\nu^{-2}\,n_{\rm e}^2\,\bar g_{\rm ff},
\label{eq:alpha_ff}
\end{equation}
where $T_{\rm{e}}$ is the electron temperature, $n_{\rm e}$ is the electron number density, and $\bar g_{\rm ff}$ is the velocity-averaged Gaunt factor of order unity.

The corresponding optical depth is obtained by integrating from the shock to the outer CSM,
\begin{equation}
\tau_{\nu,{\rm ff}} = \int_{R_{\mathrm{sh}}}^{\infty} \alpha_{\nu, \mathrm{ff}}(r) \, \mathrm{d}r.
\label{eq:tau_ff_def}
\end{equation}

Assuming the emergent intensity is approximately uniform over a spherical surface of radius $R_{\rm sh}$, the specific luminosity is
\begin{equation}
L_\nu = 4\pi^2 R_{\rm sh}^2 \, I_\nu .
\label{eq:Lnu_from_Inu}
\end{equation}
This follows from integrating the outward specific flux over the emitting area: for an isotropic intensity, the surface flux is
$F_{\nu,{\rm surf}}=\int I_\nu\cos\theta\,d\Omega=\pi I_\nu$, and thus
$L_\nu=4\pi R_{\rm sh}^2 F_{\nu,{\rm surf}}$.
Equation~(\ref{eq:Lnu_from_Inu}) implicitly assumes spherical symmetry and approximately isotropic emission in the local comoving frame. While strong relativistic effects (e.g., Doppler boosting and equal-arrival-time surfaces) can be important for ultra-relativistic outflows, the mildly relativistic velocities considered here  imply that such corrections are at most of order unity and do not affect the qualitative behavior of radio light curves.

For direct comparison with observations, we convert the model luminosity to the observed flux density via
\begin{equation}
F_{\nu,{\rm obs}}=\frac{1}{4\pi D_L^2}\frac{L_{\nu}}{1+z},
\qquad
\nu=(1+z)\nu_{\rm obs},
\label{eq:Fnu_obs_kcorr}
\end{equation}
where $D_L$ is the luminosity distance and $z$ is the source redshift.

\subsubsection{Characteristic frequencies and timescales}
The characteristic synchrotron frequency for an electron of Lorentz factor $\gamma$ scales as $\nu(\gamma)\propto B\gamma^2$. In particular, the characteristic frequencies associated with the minimum-energy and cooling electrons are
\begin{equation}
\nu_{\rm m} \equiv \frac{3q_{\rm e} B}{4\pi m_{\rm e} c}\,\gamma_{\rm m}^2, \qquad
\nu_{\rm c} \equiv \frac{3q_{\rm e} B}{4\pi m_{\rm e} c}\,\gamma_{\rm c}^2 .
\label{eq:char_freqs}
\end{equation}
In the slow-cooling regime ($\nu_{\rm m}<\nu_{\rm c}$), the optically thin synchrotron emissivity scales as
$j_{\nu,{\rm syn}}\propto \nu^{1/3}$ for $\nu<\nu_{\rm m}$ and
$j_{\nu,{\rm syn}}\propto \nu^{-(p-1)/2}$ for $\nu_{\rm m}<\nu<\nu_{\rm c}$.
The cooling frequency $\nu_{\rm c}$ corresponds to $\gamma_{\rm c}$ for which the synchrotron cooling time equals the dynamical time, and it marks the onset of cooling-induced spectral steepening at higher frequencies. In our fiducial parameter space, we typically have $\nu_{\rm obs}\ll \nu_{\rm c}$, so synchrotron radiative losses do not appreciably affect the GHz-band emission.

SSA produces a low-frequency spectral turnover.
For a power-law electron distribution, the SSA optical depth scales as
$\tau_{\nu,{\rm syn}}\propto \nu^{-(p+4)/2}$.
We define the SSA frequency $\nu_{\rm a}$ by the condition
$\tau_{\nu,{\rm syn}}(\nu_{\rm a})=1$, such that
\begin{equation}
\tau_{\nu,{\rm syn}}
=\left(\frac{\nu}{\nu_{\rm a}}\right)^{-(p+4)/2}.
\label{eq:nua_def}
\end{equation}

In the fiducial interaction model, the SSA frequency decreases approximately as
\begin{equation}
\nu_{\rm a}\simeq \nu_{{\rm a, in}}
\left(\frac{t}{t_{\rm in}}\right)^{-1},
\label{eq:nua_time}
\end{equation}
where the normalization at $t=t_{\rm in}$ is
\begin{equation}
\begin{aligned}
\nu_{{\rm a, in}}
&\simeq 1.06\times 10^{12}\,
\epsilon_{\rm B,-1}^{\frac{p+2}{2(p+4)}}
\epsilon_{\rm e,-1}^{\frac{2}{p+4}} \\
&\quad\times
\dot{M}_{-4}^{\frac{p+6}{2(p+4)}}
v_{{\rm w},3}^{-\frac{p+6}{2(p+4)}}
v_{{\rm in},-1}^{\frac{p+6}{p+4}}
R_{{\rm in},3}^{-1}\ {\rm Hz}.
\end{aligned}
\label{eq:nua_in_scaling}
\end{equation}

At a fixed observing frequency $\nu_{\rm obs}$, the SSA transparency time
$t_{\rm a}$ is defined by $\nu_{\rm a}(t_{\rm a})=\nu_{\rm obs}$, yielding
\begin{equation}
\begin{aligned}
t_{\rm a}
&= t_{\rm in}\frac{\nu_{{\rm a, in}}}{\nu_{\rm obs}} \\
&\simeq 287.3~{\rm days}\,
\epsilon_{\rm B,-1}^{\frac{p+2}{2(p+4)}}
\epsilon_{\rm e,-1}^{\frac{2}{p+4}}
\dot{M}_{-4}^{\frac{p+6}{2(p+4)}} \\
&\quad\times
v_{{\rm w},3}^{-\frac{p+6}{2(p+4)}}
v_{{\rm in},-1}^{\frac{2}{p+4}}
\nu_{{\rm obs},9}^{-1},
\end{aligned}
\label{eq:ta_nua}
\end{equation}
where $\nu_{{\rm obs},9}\equiv\nu_{\rm obs}/(10^{9}\,{\rm Hz})$.
For $t<t_{\rm a}$ the source is SSA-thick at $\nu_{\rm obs}$ and the emission is suppressed,
whereas for $t>t_{\rm a}$ the SSA opacity becomes negligible and the light curve
transitions to the optically thin synchrotron regime.
Therefore, the observed SSA peak time $t_{\rm a}$ serves as a direct diagnostic
of the CSM density normalization and shock velocity, owing to its strong
dependence on $\dot{M}$ and $v_{\rm in}$.

Similarly,  free--free absorption in the unshocked ionized CSM can impose an external low-frequency cutoff. Since $\tau_{\nu,{\rm ff}}\propto \nu^{-2}$, we define a characteristic frequency $\nu_{\rm ff}$ by
\begin{equation}
\tau_{\nu,{\rm ff}} \equiv \left(\frac{\nu}{\nu_{\rm ff}}\right)^{-2},
\label{eq:nuff_def}
\end{equation}
so that $\tau_{\nu,{\rm ff}}(\nu_{\rm ff})=1$. For a wind-like CSM, the corresponding cutoff  frequency evolves as
\begin{equation}
\nu_{\rm ff} \simeq 0.13\,T_{\rm{e}}^{-3/4}\,R_{\rm sh}^{1/2}\,n_{\rm sh}
= \nu_{{\rm ff},{\rm in}}\left(\frac{t}{t_{\rm in}}\right)^{-3m/2},
\label{eq:nuff_time}
\end{equation}
where $n_{\rm sh}\equiv n_{\rm e}(R_{\rm sh})$ is the unshocked CSM electron density at the shock radius. At $t=t_{\rm in}$ is
\begin{equation}
\begin{aligned}
\nu_{{\rm ff},{\rm in}}
&\simeq 0.13\,T_{\rm{e}}^{-3/4}\,R_{\rm in}^{1/2}\,n_{\rm in} \\
&\simeq 4.8\times 10^{11}\,{\rm Hz}
\dot{M}_{-4}\,v_{{\rm w},3}^{-1}\,T_{\rm{e},5}^{-3/4}\,
R_{{\rm in},3}^{-3/2} \, ,
\end{aligned}
\label{eq:nuff_in_scaling}
\end{equation}
where we have specialized the last expression to the steady-wind case ($s_1=2$).

The FFA transparency time $t_{\rm ff}$ at an observing frequency $\nu_{\rm obs}$ is defined by  $\nu_{\rm ff}(t_{\rm ff})=\nu_{\rm obs}$, yielding
\begin{equation}
\begin{aligned}
t_{\mathrm{ff}}
&= t_{\mathrm{in}}
\left( \frac{\nu_{\mathrm{ff,in}}}{\nu_{\mathrm{obs}}} \right)^{\frac{2}{3m}} \\
&\simeq 29 \,\mathrm{\ days}
\dot{M}_{-4}^{\frac{2}{3m}} \,
v_{\rm w, 3}^{-\frac{2}{3m}} \,
T_{\rm{e},5}^{-\frac{2}{m}} \,
R_{\mathrm{in}, 3}^{\frac{m - 1}{m}} \,
v_{\mathrm{in}, -1}^{-1} \,
\nu_{\mathrm{obs}, 9}^{-\frac{2}{3m}}  .
\end{aligned}
\label{eq:tff_nuff}
\end{equation}

For $t<t_{\rm ff}$, the emission at $\nu_{\rm obs}$ is exponentially suppressed ($F_\nu\propto e^{-\tau_{\nu,{\rm ff}}}$), whereas for $t>t_{\rm ff}$ the external screen becomes optically  thin and FFA no longer controls the light-curve rise.

\section{Results: light curves and parameter dependence} \label{sec:result}

\begin{figure*}
    \centering
    \includegraphics[width=0.9\linewidth]{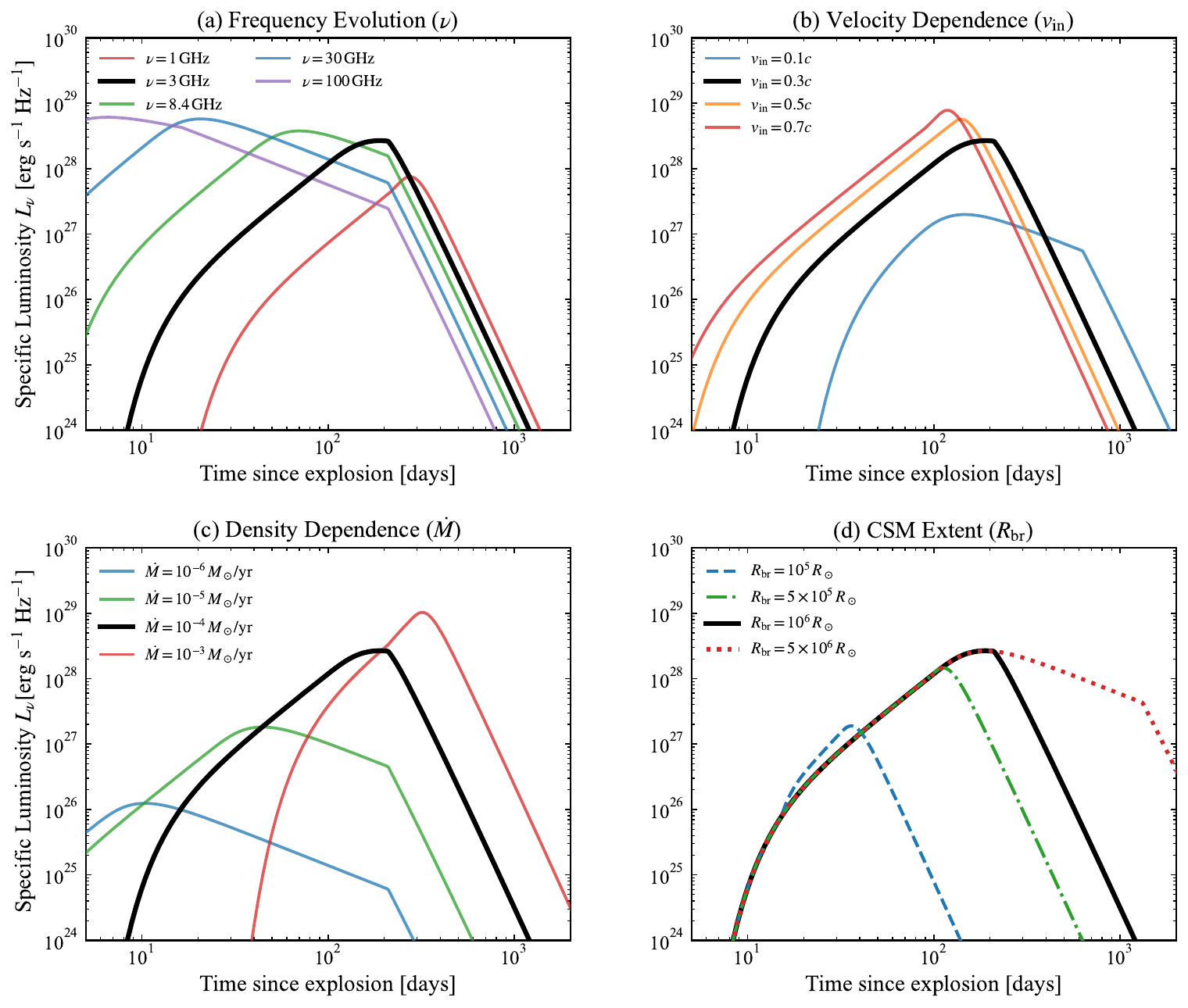}
\caption{
Radio light curves of FBOTs illustrating the dependence on key physical parameters. In all panels, the \textbf{thick black solid curve} denotes the fiducial model with $\dot{M} = 10^{-4}\,M_\odot\,{\rm yr}^{-1}$, $v_{\rm in} = 0.3c$, $R_{\rm br} = 10^6\,R_\odot$, $\nu = 3$~GHz, $\epsilon_{\rm e} = \epsilon_{\rm B} = 0.01$, and $p = 2.5$; the wind velocity is fixed at $v_{\rm w} = 1000\,{\rm km\,s^{-1}}$.
(a) Frequency dependence for $\nu = 1$, 3, 8.4, 30, and 100~GHz, with higher frequencies peaking earlier due to lower synchrotron self-absorption optical depths.
(b) Dependence on the initial shock velocity $v_{\rm in} = 0.1$--0.7$c$, where faster shocks produce brighter and more rapidly evolving emission.
(c) Dependence on the mass-loss rate $\dot{M} = 10^{-6}$--$10^{-3}\,M_\odot\,{\rm yr}^{-1}$, showing higher peak luminosities and later peak times at higher CSM densities.
(d) Dependence on the CSM break radius $R_{\rm br}$, with identical evolution before the shock reaches $R_{\rm br}$ and a rapid decline thereafter.
}
    \label{Fig:Parameter_Study}
\end{figure*}

The radio emission from FBOTs serves as a unique probe of their central engines and immediate environments. Unlike standard core-collapse supernovae, FBOTs exhibit luminous ($L_{\nu} \gtrsim 10^{28} \rm \, erg \, s^{-1} \, Hz^{-1}$) and rapidly evolving radio counterparts. To elucidate the physical origins of these extreme properties, we examine how the shock dynamics and CSM structure shape the observable radio evolution.

\subsection{Physical drivers of radio evolution} \label{sec:param_study}
The diversity of FBOT radio light curves is primarily governed by the shock velocity, the mass-loss rate, and the radial extent of the dense CSM. We explore these dependencies in Figure \ref{Fig:Parameter_Study}.

\textit{Frequency Evolution:}
The light curves show a spectral cascade expected for an expanding synchrotron source. A defining feature of FBOTs is their rapid rise to peak luminosity, most evident at millimeter frequencies. In Panel (a), at high frequencies (e.g., 100,GHz), the emission becomes optically thin to SSA and FFA at very early times ($t \lesssim 10$~days). This explains why millimeter observations are essential for probing the earliest phases of FBOT evolution, when the source is still optically thick at standard centimeter bands (e.g., 1--10,GHz).

\textit{Velocity Dependence:}
The shock velocity, $v_{\rm in}$, is the primary parameter that separates FBOTs from ordinary radio supernovae. Because the post-shock magnetic and electron energy densities scale as $\rho v_{\rm sh}^2$, the synchrotron luminosity is strongly velocity dependent. As shown in Panel (b), a typical supernova shock ($v\sim0.05c$) produces much fainter emission, whereas trans-relativistic velocities ($v_{\rm in}\sim0.1c$--$0.7c$) boost the peak spectral luminosity to $\sim10^{29},{\rm erg,s^{-1}Hz^{-1}}$, comparable to AT2018cow and CSS161010. A larger $v_{\rm sh}$ also drives faster expansion ($R_{\rm sh}\propto v_{\rm sh}t$), which increases the emitting area and accelerates the decline of the absorption optical depth, naturally yielding the rapid rises seen in FBOTs.

\textit{Density Dependence:} The mass-loss rate $\dot{M}$ governs the density of the CSM ($\rho_{\rm CSM}$). As illustrated in Panel (c), achieving the high peak luminosities characteristic of FBOTs  requires a substantial mass-loss rate ($\dot{M} \gtrsim 10^{-4}\,M_\odot\,{\rm yr}^{-1}$) to provide sufficient target material. Although such high densities typically prolong the optically thick phase by increasing SSA/FFA opacities ($\tau \propto \dot{M}^2$), the high shock velocities inherent to FBOTs counteract this effect, enabling radiation breakout on timescales of weeks rather than years.

\textit{CSM Extent:} The break radius $R_{\rm br}$ defines the outer boundary of the dense CSM shell. As demonstrated in Panel (d), the light curves follow a standard wind-interaction evolution until the shock traverses this boundary. Beyond $R_{\rm br}$, the shock enters a rarefied region with a steeper density gradient ($s_2=5$), causing the sharp luminosity turnover often observed in post-peak FBOT light curves. Identifying this ``break'' provides a direct constraint on the radial extent of the precursor mass ejection.

\subsection{Light-curve morphology}

The observed diversity of FBOT radio light curves can be naturally interpreted within a unified three-timescale framework, without invoking qualitatively distinct emission mechanisms.
Different events correspond to different orderings and separations among the free--free transparency time $t_{\rm ff}$, the synchrotron self-absorption peak time $t_{\rm a}$, and the CSM break crossing time $t_{\rm br}$, leading to a wide range of rise behaviors, peak sharpness, and post-peak decay slopes.

Events with smooth, power-law rises and well-defined peaks are expected when $t_{\rm ff}\ll t_{\rm a}\ll t_{\rm br}$.
In this regime, the early emission is dominated by SSA while the forward shock propagates through an extended dense CSM, and the light-curve peak is set by the SSA transition.
In contrast, FBOTs exhibiting delayed or flattened rises can be understood as systems with $t_{\rm ff}\gtrsim t_{\rm a}$, where external free--free absorption partially masks the intrinsic SSA evolution and shifts the apparent peak to later times.
Such cases are therefore more sensitive to the ionization state and temperature of the CSM than to the synchrotron microphysics alone.

A particularly distinctive class of FBOTs displays extremely steep post-peak declines, sometimes approaching $L_\nu\propto t^{-3}$ or steeper.
Within our framework, this behavior arises naturally when $t_{\rm br}\sim t_{\rm a}$ or $t_{\rm br}<t_{\rm a}$, implying that the forward shock exits the dense CSM shortly after, or even before, the source becomes SSA-thin.
The rapid exhaustion of interaction power in a truncated CSM then dominates the light-curve morphology, producing a sharp turnover that cannot be reproduced by a single, extended wind profile.
Steep decay indices observed in events such as CSS161010 and AT2020xnd are therefore not anomalous, but instead point to a spatially confined mass-loss history prior to explosion.

\begin{figure}
\centering
\includegraphics[width=1\linewidth]{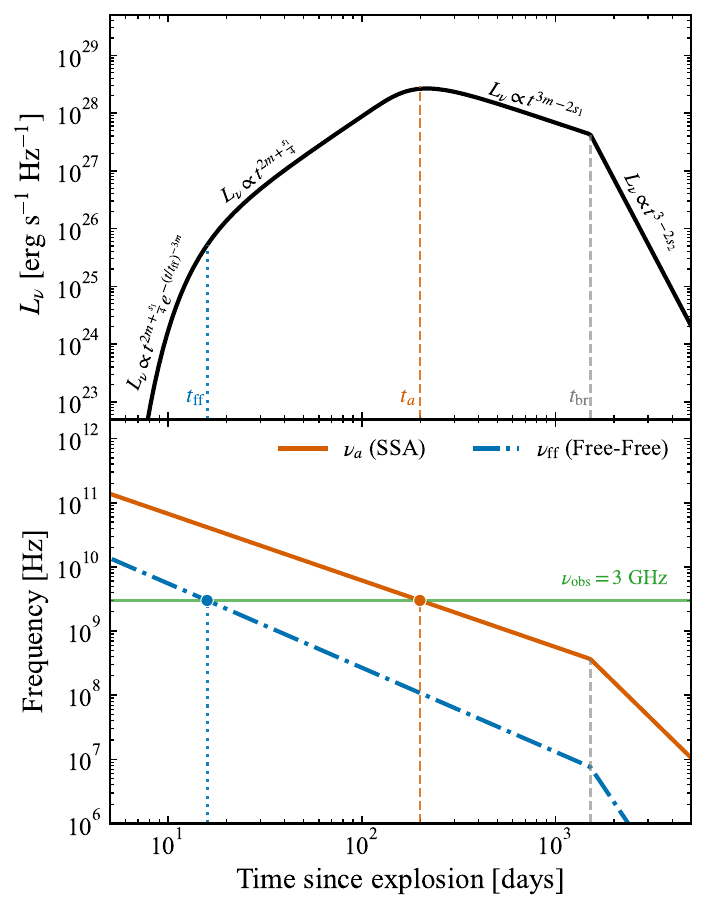}
\caption{
Fiducial radio light curve and characteristic absorption frequencies for an FBOT at $\nu_{\rm obs}=3~{\rm GHz}$.
Top panel: Specific luminosity $L_\nu$ (black) with asymptotic temporal scalings.
Bottom panel: Evolution of the synchrotron self-absorption frequency $\nu_{\rm a}$ (orange) and the free--free absorption cutoff $\nu_{\rm ff}$ (blue).
}

\label{Fig:LC_nus}
\end{figure}

Figure~\ref{Fig:LC_nus} summarizes the multi-stage radio evolution driven by synchrotron emission from the forward shock and absorption in the surrounding CSM.
At early times ($t<t_{\rm ff}$), the unshocked ionized CSM acts as an external absorbing screen, and the flux is exponentially suppressed,
$L_\nu \propto \exp(-\tau_{\nu,{\rm ff}})\simeq \exp[-(t/t_{\rm ff})^{-3m}]$,
until the free--free cutoff frequency drops below the observing band.
For $t_{\rm ff}<t<t_{\rm a}$, free--free absorption becomes negligible while the source remains SSA-thick, and the light curve rises as
$L_\nu \propto t^{2m+s_1/4}$.
The peak occurs at $t_{\rm a}$, marking the transition from SSA-thick to SSA-thin emission.

At later times ($t>t_{\rm a}$), the emission is optically thin and decays as $L_\nu \propto t^{3m-2s_1}$ as long as the shock propagates within the inner dense CSM.
A second steepening at $t_{\rm br}$ signals that the shock has crossed the CSM density break into a more rapidly declining outer profile ($s_2>s_1$); in the post-break coasting approximation, the decay asymptotes to $L_\nu \propto t^{3-2s_2}$.
The pair of observables $(t_{\rm a},\,t_{\rm br})$ therefore provides a compact and physically transparent diagnostic of both the absorption-controlled transparency time and the radial extent of the dense CSM, enabling direct constraints on $(\dot{M},\,v_{\rm in},\,R_{\rm br})$.

Overall, the radio light-curve morphology of FBOTs primarily reflects the relative timing of absorption transparency and CSM truncation, rather than requiring fundamentally different outflow properties across the population.
Multi-frequency monitoring that resolves $(t_{\rm ff},\,t_{\rm a},\,t_{\rm br})$ thus offers an observationally accessible means of classifying FBOT radio behavior and reconstructing the progenitor mass-loss history, with $t_{\rm br}$ encoding the look-back timescale of enhanced pre-explosion mass ejection.

\subsection{Comparison with SNe and GRBs} \label{sec:population}
\begin{figure}
\centering
\includegraphics[width=1\linewidth]{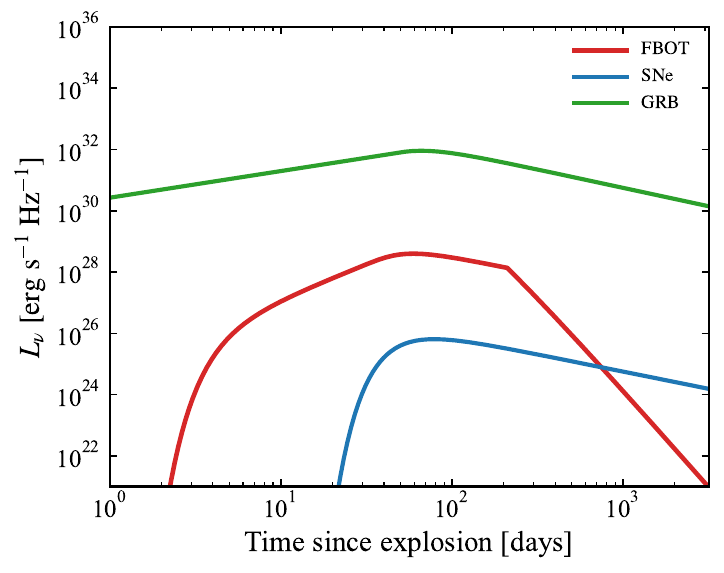}
\caption{
Representative radio light curves illustrating the characteristic luminosity and temporal evolution of ordinary supernovae (SNe; blue), fast blue optical transients (FBOTs; red), and gamma-ray burst afterglows (GRBs; green).
FBOTs occupy an intermediate regime in peak luminosity and evolution timescale, lying between the fainter, slowly evolving SNe and the more luminous, longer-lived GRB afterglows.
}

\label{Fig:Fig_FBOT_GRB_SNe_comparison}
\end{figure}

\begin{figure}
\centering
\includegraphics[width=\linewidth]{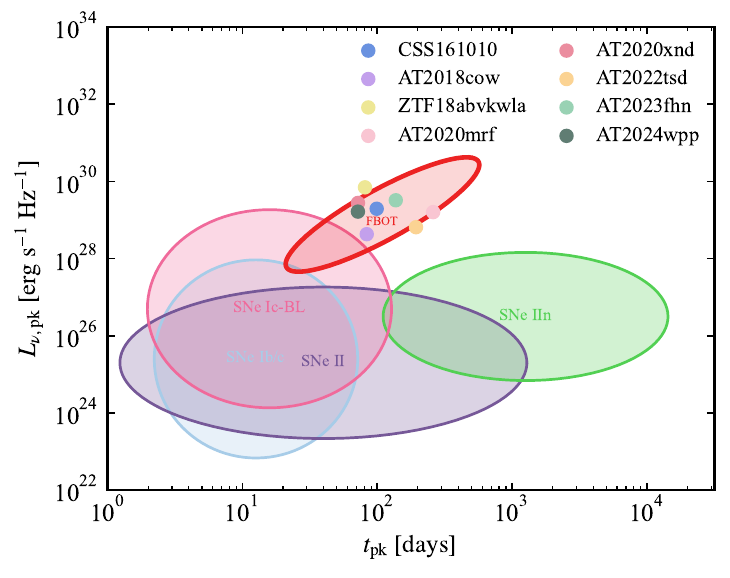}
\caption{
Peak radio properties of FBOTs compared with core-collapse supernovae.
The peak spectral luminosity, $L_{\nu,\mathrm{pk}}$, derived from radio data in the 4--10~GHz band, is shown as a function of
$(\Delta t/1~\mathrm{day})(\nu_{\mathrm{pk}}/5~\mathrm{GHz})$, where $\nu_{\mathrm{pk}}$ and
$\Delta t$ denote the peak frequency and the characteristic peak timescale, respectively.
Filled circles represent the FBOT sample.
The shaded regions for SNe Ic-BL, SNe Ib/c, SNe II, and SNe IIn are adapted from \citet{Bietenholz2021}, and approximately represent the 68\% contours
of the corresponding SN subclasses in the radio luminosity--risetime plane.
The red FBOT region schematically indicates the range covered by our model calculations
from scans over $\epsilon_B$, $\epsilon_e$, $\dot{M}$, $\nu_{\mathrm{obs}}$, and
$v_{\mathrm{in}}$.
}
\label{fig:Fig_Lpeak_tpk_comparison}
\end{figure}

To place FBOTs in a broader context, we compare their radio properties with those of other explosive transients in Figures~\ref{Fig:Fig_FBOT_GRB_SNe_comparison} and \ref{fig:Fig_Lpeak_tpk_comparison}.

Figure~\ref{Fig:Fig_FBOT_GRB_SNe_comparison} presents representative radio light curves for SN-like, FBOT-like, and GRB-like explosions, computed using the parameter sets described above. The SN-like model combines a non-relativistic ejecta velocity ($v_{\rm ej}=0.03c$) with a low-density stellar wind ($\dot{M}=10^{-4}M_\odot{\rm yr^{-1}}$). This configuration produces faint radio emission that rises slowly and peaks at late times. In contrast, the FBOT-like model features trans-relativistic ejecta ($v_{\rm ej}=0.3c$) interacting with a dense but radially confined CSM ($\dot{M}=10^{-4}M_\odot{\rm yr^{-1}}$ $R_{\rm br}=10^6R_\odot$). It generates a much brighter radio flare with a rapid rise and a well-defined peak, followed by a steep decline once the shock exits the dense CSM. The GRB-like case represents the opposite extreme. An ultra-relativistic outflow with $\Gamma_0=100$  produces radio luminosities orders of magnitude higher and evolves over much longer timescales, reflecting sustained synchrotron emission from a relativistic blast wave propagating in a low-density medium. These models demonstrate that FBOT radio emission naturally occupies an intermediate regime between ordinary core-collapse supernovae and GRB afterglows in both luminosity and temporal evolution. This intermediate behavior is primarily controlled by the outflow velocity and the radial extent and density of the surrounding medium.

Figure~\ref{fig:Fig_Lpeak_tpk_comparison} places FBOTs within the broader population of radio-emitting transients by comparing their peak spectral luminosities with their peak times. FBOTs populate a distinct region of parameter space. Their peak luminosities are systematically higher than those of SNe~II and SNe~IIn, while showing substantial overlap with SNe~Ic-BL, a class characterized by significantly higher ejecta velocities than most other supernova types. The multi-band peak measurements for individual FBOTs cluster tightly, indicating a relatively narrow range of characteristic shock properties and CSM conditions. When interpreted within an SSA-dominated interaction in a wind-like CSM, their locations are consistent with trans-relativistic shock velocities and dense but finite mass loading in the immediate environment. This placement reinforces the interpretation that FBOT radio emission is powered by compact, energetic shock interaction rather than by highly collimated relativistic jets.

\section{Application to representative FBOTs} \label{sec:Application}

We apply the forward model developed in Section \ref{sec:framework} to a set of
well-observed FBOTs with multi-frequency radio coverage. By fitting the radio
SEDs and light curves in a uniform framework, we aim to constrain the shock
dynamics and the radial structure of CSM, with a
particular focus on the origin of the unusually steep post-peak fading seen in
several events.

\subsection{Parameter setting and fitting strategy}
\label{subsec:fitting_strategy}

To mitigate parameter degeneracies, we fix the electron energy index, the CSM temperature, 
the inner CSM slope, and the shock-deceleration index to fiducial values. 
The remaining parameters, including \(v_{\rm in}\), \(s_2\), \(R_{\rm in}\), \(R_{\rm br}\), 
\(\dot{M}\), \(\epsilon_e\), and \(\epsilon_B\), are inferred from the radio data.

\begin{figure*}[htbp]
\centering
\includegraphics[width=0.90\textwidth]{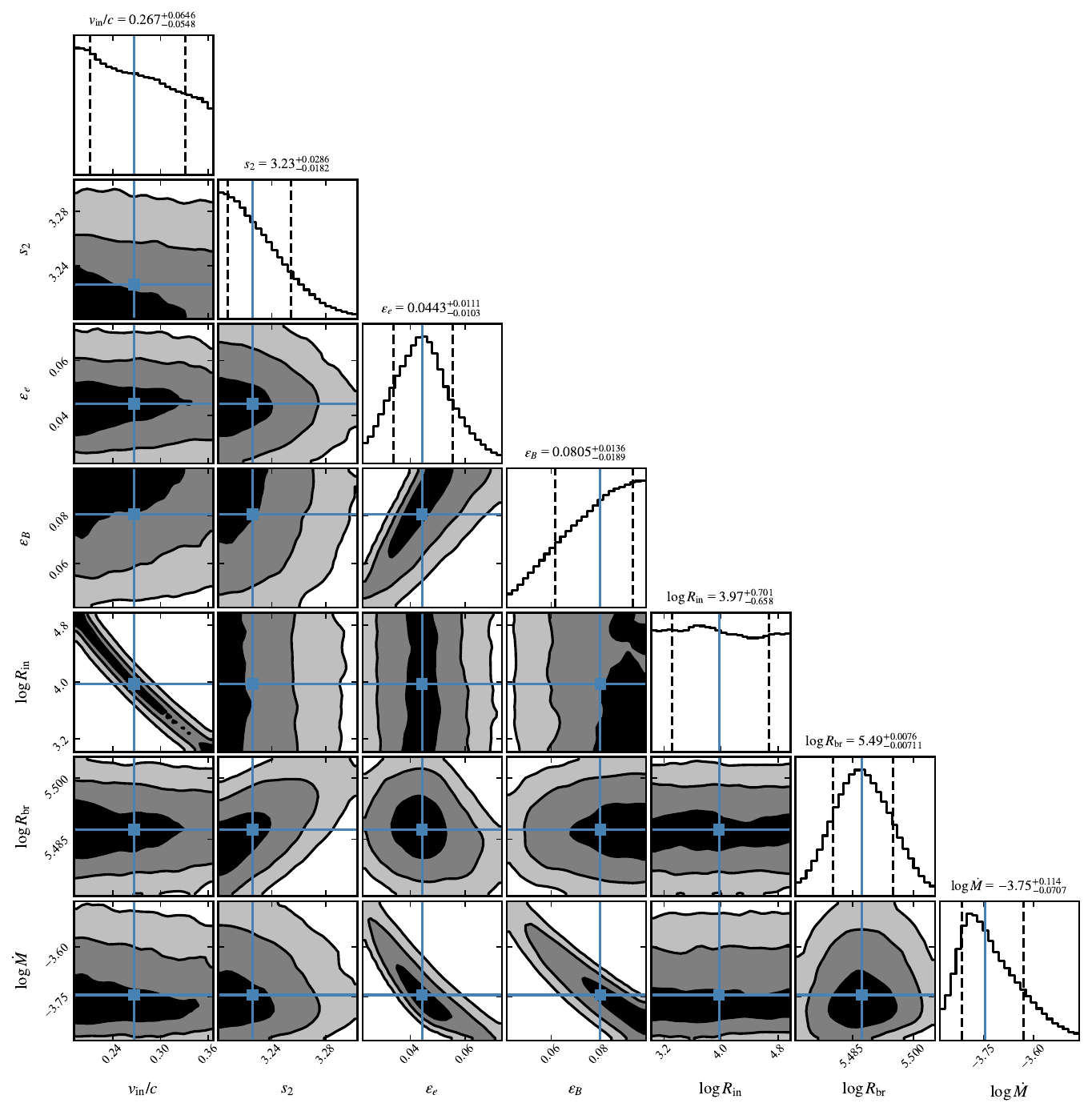}
\caption{
Corner plot of the posterior distributions for the radio light-curve model parameters of AT2024wpp. 
The diagonal panels show the marginalized one-dimensional posteriors, and the off-diagonal panels show the joint posterior distributions. 
Blue solid lines denote the median values, and black dashed lines indicate the 16th and 84th percentiles. 
}
\label{fig:corner_at2024wpp}
\end{figure*}

With these parameters fixed, we perform a Bayesian inference analysis to constrain the free parameters governing the system geometry and density structure: the outer density index $s_2$, the initial shock velocity $v_{\rm in}$, the electron and magnetic-field energy fractions $\epsilon_{\rm e}$ and $\epsilon_{\rm B}$, the effective inner radius $R_{\rm in}$, the break radius $R_{\rm br}$, and the mass-loss rate $\dot{M}$. The posterior distributions are sampled numerically, and the resulting parameter constraints are summarized in Table~\ref{tab:fbot_params_csm}, which lists the median values and associated credible intervals. As an example, the posterior corner plot for AT2024wpp is shown in \ref{fig:corner_at2024wpp}.

\subsection{Modeling the radio SEDs}
\label{subsec:sed_modeling}

\begin{figure*}
\centering
\includegraphics[width=0.95\linewidth]{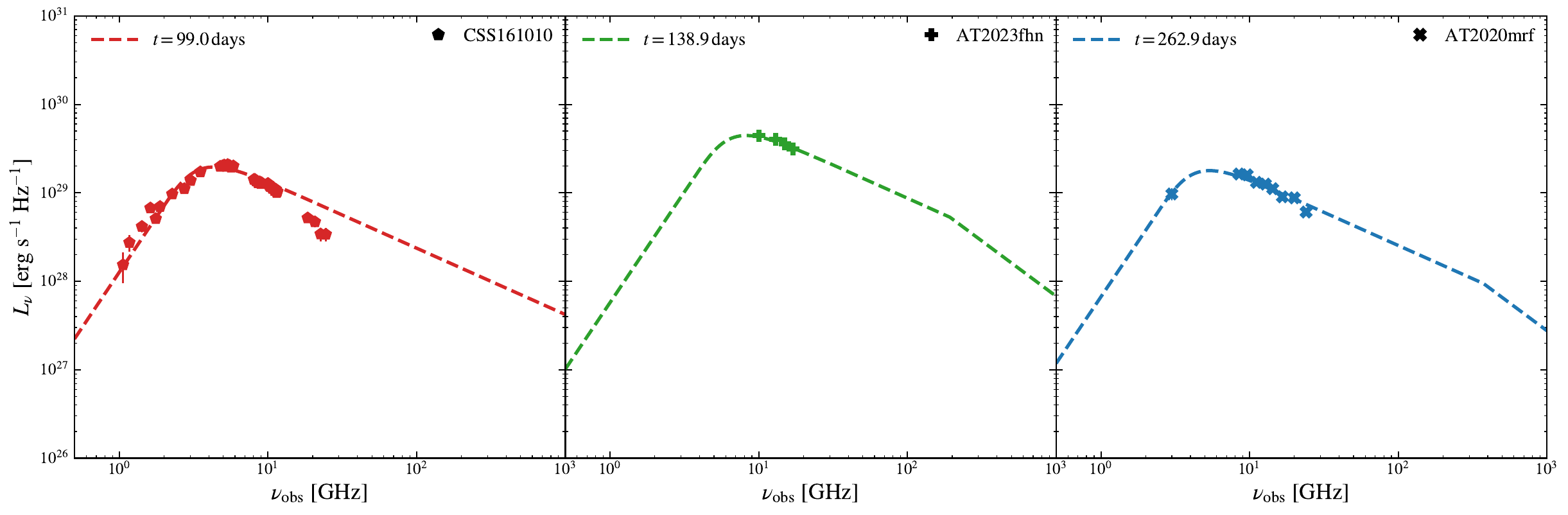}
\caption{Radio spectral luminosity density $L_\nu$ as a function of observing
frequency $\nu_{\rm obs}$ for CSS161010 ($t=99$~days), AT2023fhn ($t=138.9$~days),
and AT2020mrf ($t=262.9$~days). Filled circles show measurements with $1\sigma$
uncertainties and downward triangles denote $3\sigma$ upper limits. Dashed curves
show the best-fit synchrotron spectra, with a self-absorbed turnover at low
frequencies and an optically thin power-law decline at high frequencies.}
\label{fig:radio_sed_comp}
\end{figure*}

We first assess the model performance against broadband radio SEDs. Owing to the lack of long-term multi-frequency monitoring, only a limited number of epochs provide sufficiently complete frequency coverage for meaningful SED fitting. Figure~\ref{fig:radio_sed_comp} presents representative examples for CSS161010 ($t\simeq99$~d), AT2023fhn ($t\simeq139$~d), and AT2020mrf ($t\simeq263$~d).

In the SSA-dominated case, the optically thick branch approaches the canonical
scaling, while the optically thin spectrum follows
$L_\nu\propto \nu^{-(p-1)/2}$. With $p=2.5$, the expected optically thin index
is $-0.75$, broadly consistent with the data in
Figure~\ref{fig:radio_sed_comp}, supporting our fiducial choice of $p$.

CSS161010 shows an additional steepening at $\nu\gtrsim100$~GHz relative to a
single power-law extrapolation. Interpreting this feature solely as a cooling
break would require the synchrotron cooling frequency $\nu_{\rm c}$ to approach the
mm band at this epoch, which is difficult to accommodate in our fiducial
parameter space. A more natural interpretation is that the electron
distribution or emitting region is not adequately described by a single-zone,
single-power-law model at the highest frequencies, motivating caution when
including mm points in a minimal SSA framework.

\subsection{Modeling the light curves and steep fading}
\label{subsec:lc_modeling}

\begin{figure}
\centering
\includegraphics[width=\linewidth]{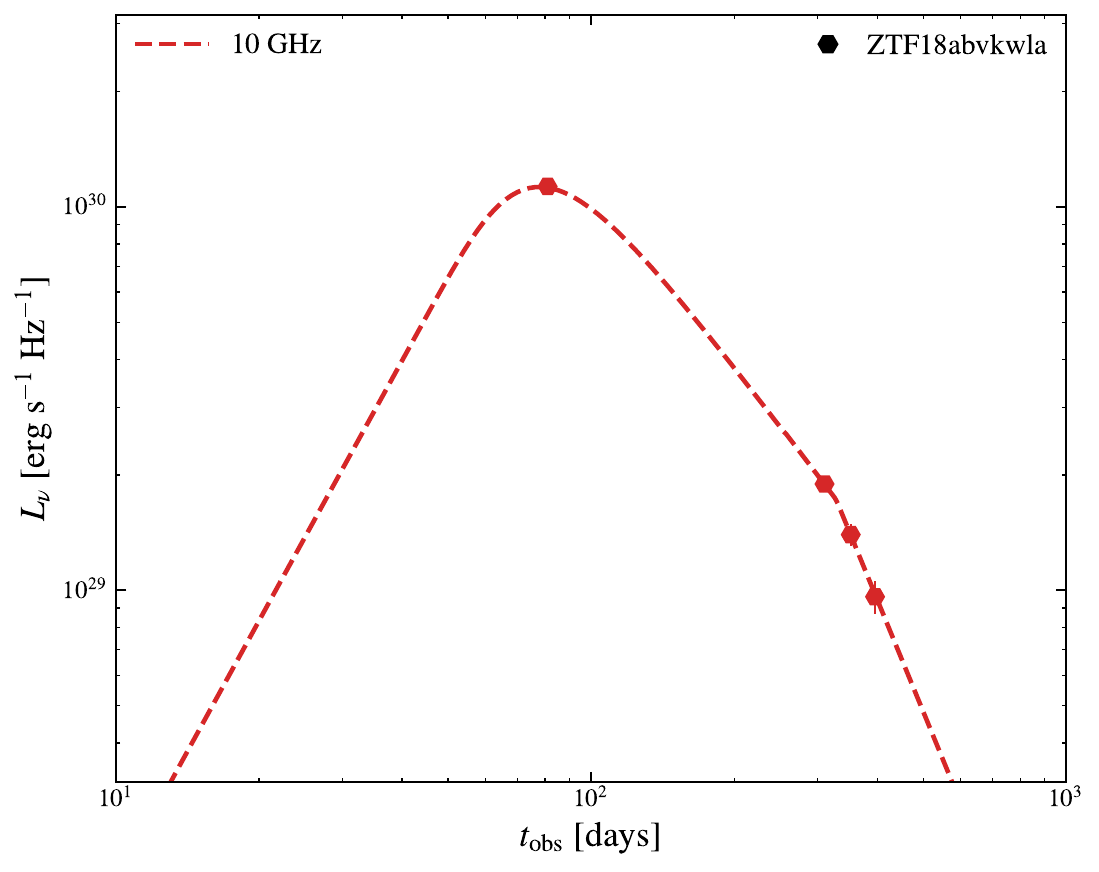}
\caption{10~GHz radio light curve of ZTF18abvkwla. Since the available data are sparse, we fit only the 10~GHz light curve.}
\label{Fig:Fig_Lc_ZTF18abvkwla}
\end{figure}

\begin{figure*}
    \centering
    \includegraphics[width=1\linewidth]{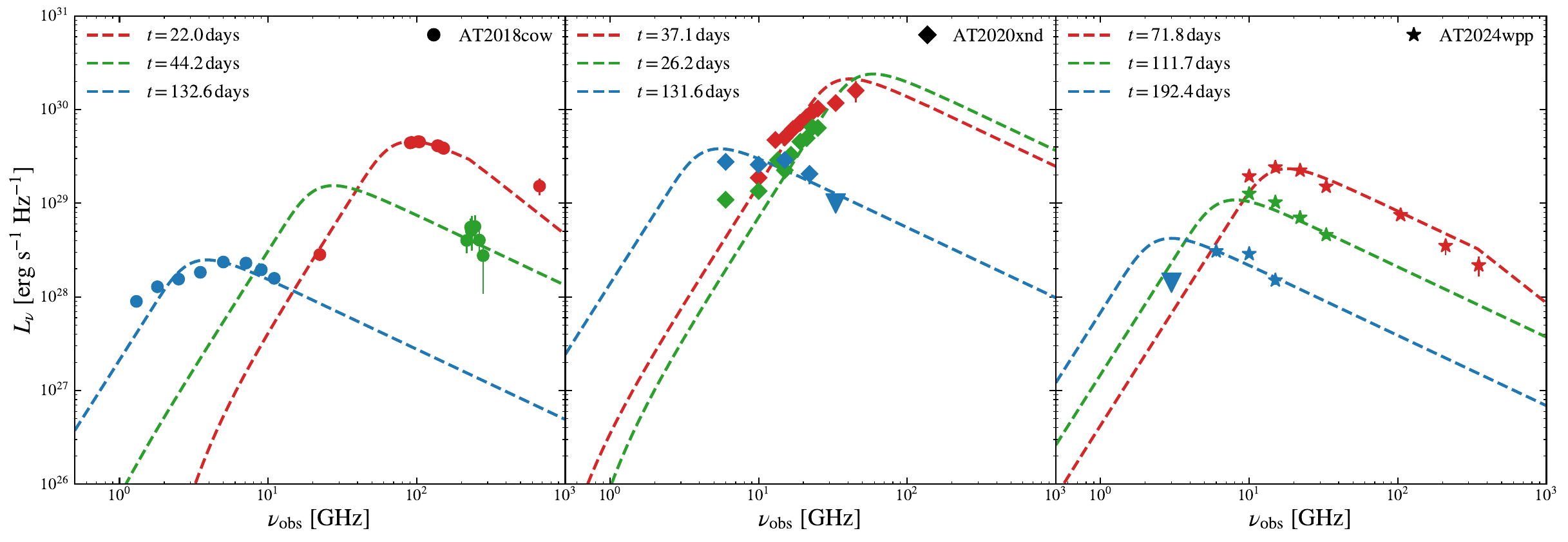}
    \includegraphics[width=1\linewidth]{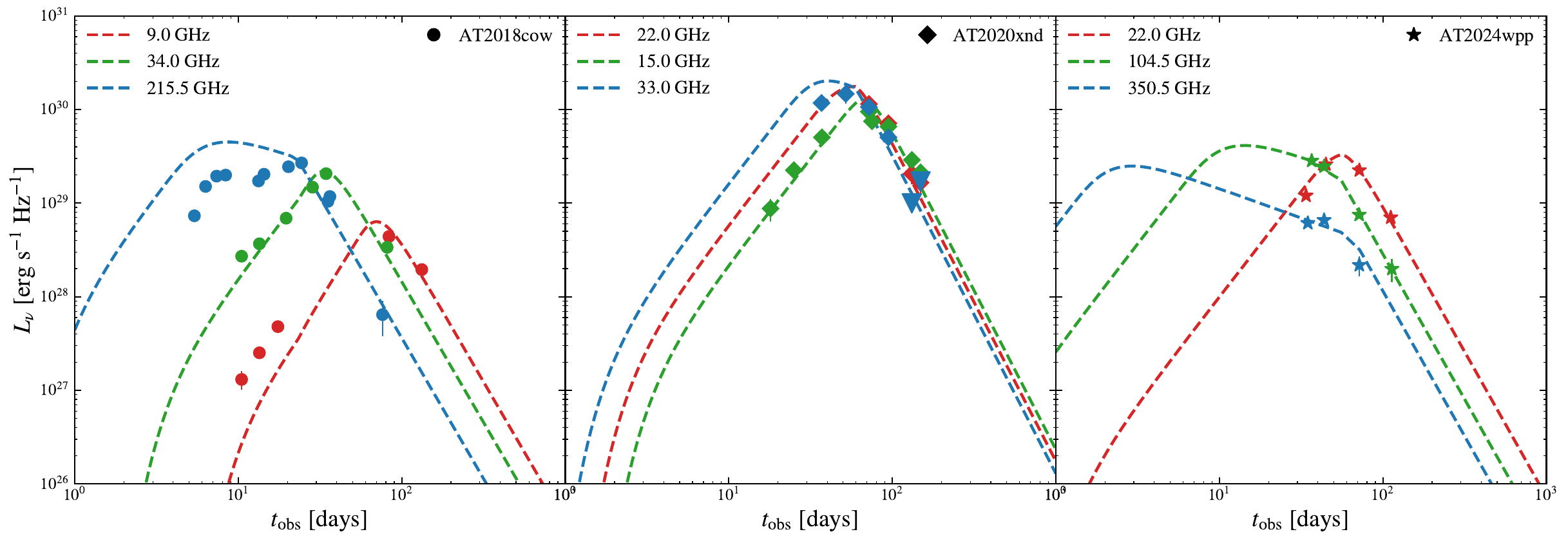}

    \caption{
    Multi-epoch fits for AT2018cow, AT2020xnd and AT2024wpp using the best-fit parameters. Upper panel: broadband radio SED at multiple epochs; dashed curves show the model spectra, and points show the observed luminosities, with non-detections plotted as inverted triangles indicating $3\sigma$ upper limits. Lower panel: multi-frequency radio light curves; dashed curves show the model predictions evaluated at the corresponding observing frequencies. Points show inferred spectral luminosities (downward triangles denote $3\sigma$ upper limits) and dashed curves show the best-fit model predictions. The rapid post-peak fading is reproduced when the forward shock reaches the CSM break radius $R_{\rm br}$ and propagates into a much steeper outer density profile.}
    \label{Fig:Three_comparison}
\end{figure*}

We next fit the multi-frequency light curves and highlight the role of the CSM
break radius $R_{\rm br}$ in shaping the post-peak behavior. The upper panel of
Figure~\ref{Fig:Three_comparison} compares the best-fit light curves for three representative
events, while the corresponding SED fits obtained using the posterior-median
parameters are also shown. Owing to the limited data available for
ZTF18abvkwla, we fit only its 10~GHz light curve, as shown in
Figure~\ref{Fig:Fig_Lc_ZTF18abvkwla}.

At early times, all three sources exhibit an absorption-regulated rise and peak
on $\lesssim 100$~day timescales, consistent with shock propagation through a
dense, wind-like inner CSM ($s_1=2$). Their late-time evolution, however, is
inconsistent with the relatively shallow decline expected for an unbroken wind
profile. In the model, the steep fading is naturally produced once the shock
traverses $R_{\rm br}$ and enters a rapidly declining outer CSM ($s_2>s_1$),
which suppresses both the synchrotron emissivity and the absorbing column.

For ZTF18abvkwla, the constraints are driven primarily by the late-time upper
limits; the preferred solution requires a transition to a steep decay at
$t\gtrsim 260$~days with $s_2\simeq 3.3$. AT2020xnd exhibits a well-sampled break
and favors $s_2\simeq 3.4$ with a steep decline at $t\gtrsim108$~days. AT2020xnd and AT2024wpp are both well observed, showing a clear rise followed by a rapid decline in their radio evolution over the monitoring window. In both cases, the shock traverses the dense CSM and reaches $R_{\rm{br}}$ in less than two months, at approximately 47 and 31 days, respectively.

AT2018cow is more complex. While the model reproduces the overall evolution of
the turnover frequency and the broad-band trends, it does not capture all
details of the multi-frequency light curves and the high-frequency SEDs
(Fig.~\ref{Fig:Three_comparison}). This is consistent with previous suggestions
that AT2018cow may involve additional physical ingredients beyond a minimal
ejecta--CSM interaction scenario (e.g., a central engine; \citealt{Ho2019}).  AT2018cow remains an outlier: its inferred characteristic timescale is much
shorter, and together with its rapid variability and unusual X-ray properties
this supports a scenario in which a central engine plays a significant role,
whereas the broader sample of radio-luminous FBOTs is consistent with shock
interaction in a dense but radially confined CSM environment.

\subsection{Physical implications: CSM extent, mass, and pre-explosion history}
\label{subsec:physical_inferences}

As summarized in Table~\ref{tab:fbot_params_csm}, FBOTs with well-sampled radio
light curves consistently favor steep outer CSM density profiles, with
$s_2 \gtrsim 3$. This indicates a rapid density decline beyond the main
interaction region and supports a radially confined CSM structure. We exclude AT~2022tsd from the population-level trends because, although a decline may be seen at some frequencies, it is not pronounced. In addition, the radio light curves remain remarkably smooth across many frequencies over the monitoring window, pointing to a likely more complex circumstellar environment than captured by our model adopted here. Meanwhile, AT~2023fhn samples only the rising phase, leaving $R_{\rm br}$ and $s_2$ poorly constrained.

For the remaining events, the inferred CSM break radii correspond to
$R_{\rm br} \sim (1$--$16)\times 10^{5}\,R_{\odot}$. Interpreted as the outer edge of
enhanced mass loading, these radii imply look-back times of order years to
decades for plausible wind velocities. The total CSM masses enclosed within
$R_{\rm br}$ are modest, $M_{\rm csm} \sim 5\times10^{-4}$--$10^{-2}\,M_\odot$,
despite the high inferred mass-loss rates. Assuming a wind velocity of
$v_w=1000~\mathrm{km\,s^{-1}}$, the onset of CSM formation is inferred to
precede the explosion by no more than $\sim 100$~yr, so that the ejecta begin
interacting with the CSM within roughly a century of the start of mass
ejection. This combination points to a brief but intense episode of mass
ejection shortly before explosion.

\begin{table*}[t]
\centering
\renewcommand{\arraystretch}{1.2}
\setlength{\tabcolsep}{4pt}
\caption{Best-fit model parameters and inferred CSM properties for the FBOT sample.}
\label{tab:fbot_params_csm}
\resizebox{\textwidth}{!}{%
\begin{tabular}{l c c c c c c c c c}
\hline\hline
\noalign{\smallskip}
& \multicolumn{7}{c}{Best-fit Parameters} & \multicolumn{2}{c}{Inferred Properties} \\
\cmidrule(lr){2-8} \cmidrule(lr){9-10}
Transient
& $v_{\mathrm{in}}/c$
& $s_2$
& $\epsilon_e$
& $\epsilon_B$
& $\log (R_{\mathrm{in}}/R_{\odot})$
& $\log (R_{\mathrm{br}}/R_{\odot})$
& $\log (\dot{M}/M_{\odot}\,\mathrm{yr}^{-1})$
& $t_{\mathrm{pre, br}}$
& $M_{\mathrm{csm}}$ \\
& & & & & & & & (yr) & ($10^{-3}\,M_{\odot}$) \\
\hline
\noalign{\smallskip}
CSS161010
& $0.494^{+0.005}_{-0.009}$
& $4.133^{+0.8079}_{-0.6721}$
& $0.109^{+0.019}_{-0.026}$
& $0.0013^{+0.0005}_{-0.0002}$
& $4.95^{+0.03}_{-0.06}$
& $6.21^{+0.02}_{-0.03}$
& $-4.01^{+0.08}_{-0.08}$
& $35.8$
& $3.30$ \\

AT2018cow
& $0.172^{+0.042}_{-0.037}$
& $3.203^{+0.0055}_{-0.0021}$
& $0.216^{+0.069}_{-0.080}$
& $0.0295^{+0.0203}_{-0.0127}$
& $3.99^{+0.66}_{-0.63}$
& $5.02^{+0.01}_{-0.02}$
& $-3.63^{+0.18}_{-0.19}$
& $2.3$
& $0.49$ \\

ZTF18abvkwla
& $0.466^{+0.025}_{-0.039}$
& $3.300^{+0.1490}_{-0.0733}$
& $0.201^{+0.079}_{-0.104}$
& $0.0061^{+0.0765}_{-0.0044}$
& $4.75^{+0.19}_{-0.31}$
& $6.04^{+0.23}_{-0.69}$
& $-3.45^{+0.33}_{-0.54}$
& $24.2$
& $8.14$ \\

AT2020mrf
& $0.231^{+0.063}_{-0.055}$
& $4.263^{+0.8099}_{-0.7402}$
& $0.084^{+0.134}_{-0.060}$
& $0.0343^{+0.1818}_{-0.0288}$
& $3.97^{+0.71}_{-0.63}$
& $6.19^{+0.21}_{-0.24}$
& $-3.56^{+0.38}_{-0.48}$
& $34.1$
& $9.35$ \\

AT2020xnd
& $0.398^{+0.010}_{-0.006}$
& $3.411^{+0.0696}_{-0.0665}$
& $0.095^{+0.023}_{-0.032}$
& $0.0051^{+0.0034}_{-0.0022}$
& $4.89^{+0.08}_{-0.14}$
& $5.73^{+0.01}_{-0.02}$
& $-3.44^{+0.16}_{-0.15}$
& $11.8$
& $3.68$ \\

AT2022tsd\textsuperscript{a}
& $0.340^{+0.085}_{-0.068}$
& $4.398^{+0.7674}_{-0.8064}$
& $0.100^{+0.137}_{-0.069}$
& $0.0290^{+0.1034}_{-0.0229}$
& $4.05^{+0.63}_{-0.65}$
& $6.17^{+0.23}_{-0.23}$
& $-4.06^{+0.40}_{-0.50}$
& $32.6$
& $2.82$ \\

AT2023fhn\textsuperscript{b}
& $0.401^{+0.071}_{-0.085}$
& $4.269^{+0.8394}_{-0.7715}$
& $0.074^{+0.133}_{-0.052}$
& $0.0138^{+0.1075}_{-0.0112}$
& $4.30^{+0.47}_{-0.68}$
& $6.12^{+0.26}_{-0.27}$
& $-3.56^{+0.35}_{-0.53}$
& $29.1$
& $7.88$ \\

AT2024wpp
& $0.267^{+0.065}_{-0.055}$
& $3.230^{+0.0286}_{-0.0182}$
& $0.0443^{+0.0111}_{-0.0103}$
& $0.0805^{+0.0136}_{-0.0189}$
& $3.97^{+0.701}_{-0.658}$
& $5.49^{+0.01}_{-0.01}$
& $-3.75^{+0.11}_{-0.07}$
& $6.8$
& $1.18$ \\
\noalign{\smallskip}
\hline
\end{tabular}%
}
\begin{flushleft}
\footnotesize
\textbf{Notes.} Columns represent: (1) Initial shock velocity; (2) outer CSM density slope; (3--4) fractions of post-shock thermal energy carried by non-thermal electrons and magnetic fields; (5--6) inner and break radii of the CSM; (7) mass-loss rate; (8) look-back time corresponding to $R_{\rm br}$, computed assuming $v_w=1000~\mathrm{km\,s^{-1}}$; (9) total CSM mass enclosed between $R_{\rm in}$ and $R_{\rm br}$, computed for $s_1=2$ and $v_w=1000~\mathrm{km\,s^{-1}}$.\\
\textsuperscript{a} Parameters for AT2022tsd are all uncertain because its radio evolution is comparatively smooth over the monitoring window.\\
\textsuperscript{b} Parameters for AT2023fhn are constrained primarily by the rising phase; $R_{\rm br}$ and $s_2$ remain more model-dependent because of the lack of post-peak data.
\end{flushleft}
\end{table*}

\section{Discussions and conclusions}
\label{sec:discussion}

In this work, we have investigated the radio emission of FBOTs using a self-consistent forward-shock synchrotron model interacting with a radially confined CSM. By fitting multi-frequency radio light curves and spectra within a single physical framework, our approach moves beyond traditional epoch-by-epoch analyses to provide a coherent description of both the temporal and spectral evolution.

A key finding of this study is that the observed diversity of FBOT radio light curves can be explained by shock interaction with a dense but radially confined CSM.
In this picture, the rise and peak of the radio emission are regulated by absorption effects, while the post-peak evolution is governed primarily by the radial extent of the dense CSM.
In particular, the exceptionally steep declines observed in several well-sampled FBOTs arise naturally once the forward shock exits the dense inner region and propagates into a lower-density environment.
Such behavior is difficult to reproduce with an extended steady wind, but follows straightforwardly from a scenario in which the progenitor undergoes a brief episode of enhanced mass loss prior to explosion, producing a finite interaction region.
The characteristic timescales associated with absorption transparency and shock propagation provide a convenient phenomenological description of this evolution, but the underlying physical driver is the confined nature of the CSM.

The inferred shock velocities cluster in the trans-relativistic regime, with $v_{\rm in}/c \simeq 0.17$--$0.5$ for the majority of the sample, significantly exceeding those of ordinary core-collapse supernovae while remaining well below ultra-relativistic GRB outflows.  It is worth noting that not all radio-luminous FBOTs follow this evolutionary path. In particular, AT2019ijn exhibits an exceptionally luminous and long-lived radio counterpart, peaking at $\sim600$ days after the optical discovery, with properties more reminiscent of off-axis jetted events than of confined CSM interaction (H.-C. Ding et al. 2026). Such events likely represent a distinct, engine-dominated channel within the broader FBOT population.

Such velocities, combined with dense CSM characterized by mass-loading rates of $\dot{M} \sim 10^{-4}$--$10^{-3.3}\,M_\odot\,\mathrm{yr^{-1}}$, naturally give rise to luminous radio emission with peak spectral luminosities of $L_\nu \sim 10^{28}$--$10^{30}\,\mathrm{erg\,s^{-1}\,Hz^{-1}}$ on timescales of weeks to months. 
Despite these high mass-loading rates, the inferred CSM is radially confined, with break radii $R_{\rm br} \sim (1$--$16)\times10^{5}\,R_\odot$, corresponding to pre-explosion mass-loss episodes occurring only $\sim 2$--$36$~yr before core collapse for typical wind velocities. 
As a result, the total CSM masses remain modest, $M_{\rm csm} \sim 5\times10^{-4}$--$10^{-2}\,M_\odot$, favoring a scenario in which FBOT progenitors undergo brief but intense mass-ejection episodes rather than long-lived steady winds, and demonstrating that extreme radio luminosities can be achieved without invoking highly collimated relativistic jets.

Several simplifying assumptions adopted in this work merit further discussion. 
First, we have neglected the contribution of secondary electrons produced in hadronic interactions. 
In sufficiently dense environments, proton--proton collisions can inject an additional population of relativistic electrons that may contribute to the synchrotron emission, particularly at late times or low frequencies \citep{Petropoulou2016}. 
While the good agreement between our model and the observed radio light curves suggests that primary shock-accelerated electrons dominate in the events considered here, secondary electrons could become important in systems with higher densities or more extended interaction phases and should be incorporated in future extensions of the model. 
Second, our synchrotron calculations assume that the emission is dominated by a non-thermal power-law electron population. 
However, recent studies have suggested that the unusually steep optically thin spectra observed in some FBOTs may arise from synchrotron emission by a thermal or quasi-thermal electron population \citep{Margalit2021}. 
If thermal electrons contribute significantly, the inferred microphysical parameters and spectral interpretations may differ from those obtained under a purely non-thermal assumption, motivating hybrid models that treat thermal and non-thermal electrons self-consistently. 
Finally, we have assumed a spherically symmetric CSM. 
Observational and theoretical studies of prototypical events such as AT2018cow indicate that strong asphericity may play an important role, with the radio and X-ray emission potentially powered by shocks propagating through an anisotropic or equatorially concentrated CSM \citep{Margutti2019,Govreen-Segal2026}. 
While our model captures the global radio evolution of most FBOTs, extending it to multi-dimensional CSM structures will be essential for understanding extreme or highly variable events.

Finally, although this paper has focused primarily on radio properties, FBOTs are intrinsically multi-wavelength phenomena. Recent multi-wavelength analyses of luminous FBOT samples demonstrate diverse X-ray and radio behaviors correlated with optical evolution, highlighting the need for coordinated observations across the electromagnetic spectrum to fully characterize these transients \citep{Sevilla2026}. Combining radio constraints with optical light curves and X-ray observations provides a powerful means of disentangling their physical origin. In particular, the joint consideration of shock velocities, CSM structure, thermal optical emission, and high-energy behavior can help distinguish between shock-powered explosions and scenarios involving central engines, ultimately establishing a unified physical picture of the FBOT progenitor systems.

\begin{acknowledgements}
We thank Prof. Xinwen Shu for helpful discussions that improved this study.
This work was supported by the National Natural Science Foundation of China
(grant Nos. 12303047 and 12393811), the Natural Science Foundation of Hubei Province
(grant No. 2023AFB321), and the National Key R\&D Program of China
(grant No. 2021YFA0718500).
\end{acknowledgements}

\bibliographystyle{aa}
\bibliography{sample701}

\begin{thebibliography}{37}
\expandafter\ifx\csname natexlab\endcsname\relax\def\natexlab#1{#1}\fi

\bibitem[{{Bietenholz} {et~al.}(2021){Bietenholz}, {Bartel}, {Argo}, {Dua}, {Ryder}, \& {Soderberg}}]{Bietenholz2021}
{Bietenholz}, M.~F., {Bartel}, N., {Argo}, M., {et~al.} 2021, \apj, 908, 75

\bibitem[{{Bright} {et~al.}(2022){Bright}, {Margutti}, {Matthews}, {Brethauer}, {Coppejans}, {Wieringa}, {Metzger}, {DeMarchi}, {Laskar}, {Romero}, {Alexander}, {Horesh}, {Migliori}, {Chornock}, {Berger}, {Bietenholz}, {Devlin}, {Dicker}, {Jacobson-Gal{\'a}n}, {Mason}, {Milisavljevic}, {Motta}, {Mroczkowski}, {Ramirez-Ruiz}, {Rhodes}, {Sarazin}, {Sfaradi}, \& {Sievers}}]{Bright2022}
{Bright}, J.~S., {Margutti}, R., {Matthews}, D., {et~al.} 2022, \apj, 926, 112

\bibitem[{{Chevalier}(1982)}]{Chevalier1982}
{Chevalier}, R.~A. 1982, \apj, 258, 790

\bibitem[{{Chevalier}(1998)}]{Chevalier1998}
{Chevalier}, R.~A. 1998, \apj, 499, 810

\bibitem[{{Chevalier} \& {Fransson}(2006)}]{Chevalier2006}
{Chevalier}, R.~A. \& {Fransson}, C. 2006, \apj, 651, 381

\bibitem[{Chrimes {et~al.}(2024)Chrimes, Laskar, Margutti, Lunnan, Alexander, Chornock, \& Metzger}]{Chrimes2024_AT2023fhn}
Chrimes, A.~A., Laskar, T., Margutti, R., {et~al.} 2024, Astronomy \& Astrophysics, 691, A329

\bibitem[{{Coppejans} {et~al.}(2020){Coppejans}, {Margutti}, {Terreran}, {Nayana}, {Coughlin}, {Laskar}, {Alexander}, {Bietenholz}, {Caprioli}, {Chandra}, {Drout}, {Frederiks}, {Frohmaier}, {Hurley}, {Kochanek}, {MacLeod}, {Meisner}, {Nugent}, {Ridnaia}, {Sand}, {Svinkin}, {Ward}, {Yang}, {Baldeschi}, {Chilingarian}, {Dong}, {Esquivia}, {Fong}, {Guidorzi}, {Lundqvist}, {Milisavljevic}, {Paterson}, {Reichart}, {Shappee}, {Stroh}, {Valenti}, {Zauderer}, \& {Zhang}}]{Coppejans2020}
{Coppejans}, D.~L., {Margutti}, R., {Terreran}, G., {et~al.} 2020, \apjl, 895, L23

\bibitem[{{Drout} {et~al.}(2014){Drout}, {Chornock}, {Soderberg}, {Sanders}, {McKinnon}, {Rest}, {Foley}, {Milisavljevic}, {Margutti}, {Berger}, {Calkins}, {Fong}, {Gezari}, {Huber}, {Kankare}, {Kirshner}, {Leibler}, {Lunnan}, {Mattila}, {Marion}, {Narayan}, {Riess}, {Roth}, {Scolnic}, {Smartt}, {Tonry}, {Burgett}, {Chambers}, {Hodapp}, {Jedicke}, {Kaiser}, {Magnier}, {Metcalfe}, {Morgan}, {Price}, \& {Waters}}]{Drout2014}
{Drout}, M.~R., {Chornock}, R., {Soderberg}, A.~M., {et~al.} 2014, \apj, 794, 23

\bibitem[{{Govreen-Segal} {et~al.}(2026){Govreen-Segal}, {Nakar}, {Hotokezaka}, {Irwin}, \& {Quataert}}]{Govreen-Segal2026}
{Govreen-Segal}, T., {Nakar}, E., {Hotokezaka}, K., {Irwin}, C.~M., \& {Quataert}, E. 2026, arXiv e-prints, arXiv:2601.18887

\bibitem[{Ho {et~al.}(2023)Ho, Perley, Chen, {et~al.}}]{Ho2023_AT2022tsd_Nature}
Ho, A. Y.~Q., Perley, D.~A., Chen, P., {et~al.} 2023, Nature, 623, 927

\bibitem[{{Ho} {et~al.}(2020){Ho}, {Perley}, {Kulkarni}, {Dong}, {De}, {Chandra}, {Andreoni}, {Bellm}, {Burdge}, {Coughlin}, {Dekany}, {Feeney}, {Frederiks}, {Fremling}, {Golkhou}, {Graham}, {Hale}, {Helou}, {Horesh}, {Kasliwal}, {Laher}, {Masci}, {Miller}, {Porter}, {Ridnaia}, {Rusholme}, {Shupe}, {Soumagnac}, \& {Svinkin}}]{Ho2020}
{Ho}, A. Y.~Q., {Perley}, D.~A., {Kulkarni}, S.~R., {et~al.} 2020, \apj, 895, 49

\bibitem[{{Ho} {et~al.}(2019){Ho}, {Phinney}, {Ravi}, {Kulkarni}, {Petitpas}, {Emonts}, {Bhalerao}, {Blundell}, {Cenko}, {Dobie}, {Howie}, {Kamraj}, {Kasliwal}, {Murphy}, {Perley}, {Sridharan}, \& {Yoon}}]{Ho2019}
{Ho}, A. Y.~Q., {Phinney}, E.~S., {Ravi}, V., {et~al.} 2019, \apj, 871, 73

\bibitem[{Ho {et~al.}(2022)}]{Ho2022_AT2020xnd}
Ho, A. Y.~Q. {et~al.} 2022, ApJ, 932, 116

\bibitem[{{Kashiyama} \& {Quataert}(2015)}]{Kashiyama2015}
{Kashiyama}, K. \& {Quataert}, E. 2015, \mnras, 451, 2656

\bibitem[{{Katz} {et~al.}(2010){Katz}, {Budnik}, \& {Waxman}}]{Katz2010}
{Katz}, B., {Budnik}, R., \& {Waxman}, E. 2010, \apj, 716, 781

\bibitem[{{Kremer} {et~al.}(2021){Kremer}, {Lu}, {Piro}, {Chatterjee}, {Rasio}, \& {Ye}}]{Kremer2021}
{Kremer}, K., {Lu}, W., {Piro}, A.~L., {et~al.} 2021, \apj, 911, 104

\bibitem[{{Leung} {et~al.}(2020){Leung}, {Blinnikov}, {Nomoto}, {Baklanov}, {Sorokina}, \& {Tolstov}}]{Leung2020}
{Leung}, S.-C., {Blinnikov}, S., {Nomoto}, K., {et~al.} 2020, \apj, 903, 66

\bibitem[{{Liu} {et~al.}(2023){Liu}, {Liu}, {Yu}, \& {Zhu}}]{Liu2023}
{Liu}, J.-F., {Liu}, L.-D., {Yu}, Y.-W., \& {Zhu}, J.-P. 2023, \apj, 946, 35

\bibitem[{{Liu} {et~al.}(2022){Liu}, {Zhu}, {Liu}, {Yu}, \& {Zhang}}]{Liu2022}
{Liu}, J.-F., {Zhu}, J.-P., {Liu}, L.-D., {Yu}, Y.-W., \& {Zhang}, B. 2022, \apjl, 935, L34

\bibitem[{{Margalit} \& {Quataert}(2021)}]{Margalit2021}
{Margalit}, B. \& {Quataert}, E. 2021, \apjl, 923, L14

\bibitem[{{Margutti} {et~al.}(2019){Margutti}, {Metzger}, {Chornock}, {Vurm}, {Roth}, {Grefenstette}, {Savchenko}, {Cartier}, {Steiner}, {Terreran}, {Margalit}, {Migliori}, {Milisavljevic}, {Alexander}, {Bietenholz}, {Blanchard}, {Bozzo}, {Brethauer}, {Chilingarian}, {Coppejans}, {Ducci}, {Ferrigno}, {Fong}, {G{\"o}tz}, {Guidorzi}, {Hajela}, {Hurley}, {Kuulkers}, {Laurent}, {Mereghetti}, {Nicholl}, {Patnaude}, {Ubertini}, {Banovetz}, {Bartel}, {Berger}, {Coughlin}, {Eftekhari}, {Frederiks}, {Kozlova}, {Laskar}, {Svinkin}, {Drout}, {MacFadyen}, \& {Paterson}}]{Margutti2019}
{Margutti}, R., {Metzger}, B.~D., {Chornock}, R., {et~al.} 2019, \apj, 872, 18

\bibitem[{{Moriya} {et~al.}(2013){Moriya}, {Maeda}, {Taddia}, {Sollerman}, {Blinnikov}, \& {Sorokina}}]{Moriya2013}
{Moriya}, T.~J., {Maeda}, K., {Taddia}, F., {et~al.} 2013, \mnras, 435, 1520

\bibitem[{{Murase} {et~al.}(2011){Murase}, {Thompson}, {Lacki}, \& {Beacom}}]{Murase2011}
{Murase}, K., {Thompson}, T.~A., {Lacki}, B.~C., \& {Beacom}, J.~F. 2011, \prd, 84, 043003

\bibitem[{{Nayana} \& {Chandra}(2021)}]{Nayana2021_AT2018cow}
{Nayana}, A.~J. \& {Chandra}, P. 2021, \apjl, 912, L9

\bibitem[{Nayana {et~al.}(2025)Nayana, Margutti, Wiston, Laskar, Migliori, Chornock, Galvin, LeBaron, Hajela, Christy, Sfaradi, Tsuna, Aspegren, De~Colle, {et~al.}}]{Nayana2025_AT2024wpp}
Nayana, A.~J., Margutti, R., Wiston, E., {et~al.} 2025, ApJ Letters, 993, L6

\bibitem[{{Pellegrino} {et~al.}(2022){Pellegrino}, {Howell}, {Vink{\'o}}, {Gangopadhyay}, {Xiang}, {Arcavi}, {Brown}, {Burke}, {Hiramatsu}, {Hosseinzadeh}, {Li}, {McCully}, {Misra}, {Newsome}, {Gonzalez}, {Pritchard}, {Valenti}, {Wang}, \& {Zhang}}]{Pellegrino2022}
{Pellegrino}, C., {Howell}, D.~A., {Vink{\'o}}, J., {et~al.} 2022, \apj, 926, 125

\bibitem[{{Perley} {et~al.}(2026){Perley}, {Ho}, {et~al.}}]{Perley2026_AT2024wpp}
{Perley}, D.~A., {Ho}, A. Y.~Q., {et~al.} 2026, arXiv e-prints, arXiv:2601.03337

\bibitem[{{Perley} {et~al.}(2019){Perley}, {Mazzali}, {Yan}, {Cenko}, {Gezari}, {Taggart}, {Blagorodnova}, {Fremling}, {Mockler}, {Singh}, {Tominaga}, {Tanaka}, {Watson}, {Ahumada}, {Anupama}, {Ashall}, {Becerra}, {Bersier}, {Bhalerao}, {Bloom}, {Butler}, {Copperwheat}, {Coughlin}, {De}, {Drake}, {Duev}, {Frederick}, {Gonz{\'a}lez}, {Goobar}, {Heida}, {Ho}, {Horst}, {Hung}, {Itoh}, {Jencson}, {Kasliwal}, {Kawai}, {Khanam}, {Kulkarni}, {Kumar}, {Kumar}, {Kutyrev}, {Lee}, {Maeda}, {Mahabal}, {Murata}, {Neill}, {Ngeow}, {Penprase}, {Pian}, {Quimby}, {Ramirez-Ruiz}, {Richer}, {Rom{\'a}n-Z{\'u}{\~n}iga}, {Sahu}, {Srivastav}, {Socia}, {Sollerman}, {Tachibana}, {Taddia}, {Tinyanont}, {Troja}, {Ward}, {Wee}, \& {Yu}}]{Perley2019}
{Perley}, D.~A., {Mazzali}, P.~A., {Yan}, L., {et~al.} 2019, \mnras, 484, 1031

\bibitem[{{Petropoulou} {et~al.}(2016){Petropoulou}, {Kamble}, \& {Sironi}}]{Petropoulou2016}
{Petropoulou}, M., {Kamble}, A., \& {Sironi}, L. 2016, \mnras, 460, 44

\bibitem[{{Pursiainen} {et~al.}(2018){Pursiainen}, {Childress}, {Smith}, {Prajs}, {Sullivan}, {Davis}, {Foley}, {Asorey}, {Calcino}, {Carollo}, {Curtin}, {D'Andrea}, {Glazebrook}, {Gutierrez}, {Hinton}, {Hoormann}, {Inserra}, {Kessler}, {King}, {Kuehn}, {Lewis}, {Lidman}, {Macaulay}, {M{\"o}ller}, {Nichol}, {Sako}, {Sommer}, {Swann}, {Tucker}, {Uddin}, {Wiseman}, {Zhang}, {Abbott}, {Abdalla}, {Allam}, {Annis}, {Avila}, {Brooks}, {Buckley-Geer}, {Burke}, {Carnero Rosell}, {Carrasco Kind}, {Carretero}, {Castander}, {Cunha}, {Davis}, {De Vicente}, {Diehl}, {Doel}, {Eifler}, {Flaugher}, {Fosalba}, {Frieman}, {Garc{\'\i}a-Bellido}, {Gruen}, {Gruendl}, {Gutierrez}, {Hartley}, {Hollowood}, {Honscheid}, {James}, {Jeltema}, {Kuropatkin}, {Li}, {Lima}, {Maia}, {Martini}, {Menanteau}, {Ogando}, {Plazas}, {Roodman}, {Sanchez}, {Scarpine}, {Schindler}, {Smith}, {Soares-Santos}, {Sobreira}, {Suchyta}, {Swanson}, {Tarle}, {Tucker}, {Walker}, \& {DES Collaboration}}]{Pursiainen2018}
{Pursiainen}, M., {Childress}, M., {Smith}, M., {et~al.} 2018, \mnras, 481, 894

\bibitem[{{Rybicki} \& {Lightman}(1979)}]{Rybicki1979}
{Rybicki}, G.~B. \& {Lightman}, A.~P. 1979, {Radiative processes in astrophysics } (New York: Wiley)

\bibitem[{{Sevilla} {et~al.}(2026){Sevilla}, {Ho}, {Nayana A.}, {Schulze}, {Perley}, {Bremer}, {Andreoni}, {Altunin}, {Brink}, {Chandra}, {Chen}, {Chrimes}, {Coughlin}, {Das}, {Drake}, {Filippenko}, {Fremling}, {Freeburn}, {Gal Yam}, {Gerhart}, {Graham}, {Helou}, {Hinds}, {Johnson}, {Kasliwal}, {Kumar}, {Laher}, {LeBaron}, {Li}, {Liu}, {Margalit}, {Qin}, {Rehemtulla}, {Risin}, {Rose}, {Roy}, {Rusholme}, {Schroeder}, {Sollerman}, {Wang}, {Wise}, {Yang}, {Yao}, \& {Zheng}}]{Sevilla2026}
{Sevilla}, C., {Ho}, A. Y.~Q., {Nayana A.}, J., {et~al.} 2026, arXiv e-prints, arXiv:2601.18926

\bibitem[{{Sturner} {et~al.}(1997){Sturner}, {Skibo}, {Dermer}, \& {Mattox}}]{Sturner1997}
{Sturner}, S.~J., {Skibo}, J.~G., {Dermer}, C.~D., \& {Mattox}, J.~R. 1997, \apj, 490, 619

\bibitem[{{Weiler} {et~al.}(2002){Weiler}, {Panagia}, {Montes}, \& {Sramek}}]{Weiler2002}
{Weiler}, K.~W., {Panagia}, N., {Montes}, M.~J., \& {Sramek}, R.~A. 2002, \araa, 40, 387

\bibitem[{{Xiang} {et~al.}(2021){Xiang}, {Wang}, {Lin}, {Mo}, {Lin}, {Burke}, {Hiramatsu}, {Hosseinzadeh}, {Howell}, {McCully}, {Valenti}, {Vink{\'o}}, {Wheeler}, {Ehgamberdiev}, {Mirzaqulov}, {B{\'o}di}, {Bogn{\'a}r}, {Cseh}, {Hanyecz}, {Ign{\'a}cz}, {Kalup}, {K{\"o}nyves-T{\'o}th}, {Kriskovics}, {Ordasi}, {P{\'a}l}, {S{\'a}rneczky}, {Seli}, {Szak{\'a}ts}, {Arranz-Heras}, {Benavides-Palencia}, {Cejudo-Mart{\'\i}nez}, {De la Fuente-Fern{\'a}ndez}, {Escart{\'\i}n-P{\'e}rez}, {Garc{\'\i}a-De la Cuesta}, {Gonz{\'a}lez-Carballo}, {Gonz{\'a}lez-Farf{\'a}n}, {Lim{\'o}n-Mart{\'\i}nez}, {Mantero}, {Naves-Nogu{\'e}s}, {Morales-Aimar}, {Ru{\'\i}z-Ru{\'\i}z}, {Sold{\'a}n-Alfaro}, {Valero-P{\'e}rez}, {Violat-Bordonau}, {Zhang}, {Zhang}, {Li}, {Chen}, {Sai}, \& {Li}}]{Xiang2021}
{Xiang}, D., {Wang}, X., {Lin}, W., {et~al.} 2021, \apj, 910, 42

\bibitem[{{Yao} {et~al.}(2022){Yao}, {Ho}, {Medvedev}, {Nayana}, {Perley}, {Kulkarni}, {Chandra}, {Sazonov}, {Gilfanov}, {Khorunzhev}, {Khatami}, \& {Sunyaev}}]{Yao2022_AT2020mrf}
{Yao}, Y., {Ho}, A. Y.~Q., {Medvedev}, P., {et~al.} 2022, \apj, 934, 104

\bibitem[{{Yu} {et~al.}(2015){Yu}, {Li}, \& {Dai}}]{Yu2015}
{Yu}, Y.-W., {Li}, S.-Z., \& {Dai}, Z.-G. 2015, \apjl, 806, L6

\end{thebibliography}

% \clearpage
% \twocolumn

% \appendix

% \section{Posterior distribution for AT2024wpp}
% \label{app:corner_at2024wpp}

% Figure~\ref{fig:corner_at2024wpp} shows the posterior distributions of the main
% model parameters for AT2024wpp obtained from the Bayesian inference analysis.
% The contours illustrate the parameter covariances, while the marginalized
% one-dimensional distributions indicate the constraints reported in
% Table~\ref{tab:fbot_params_csm}.
% \onecolumn
% \begin{figure}[!htbp]
% \centering
% \includegraphics[width=0.70\textwidth]{Fig_AT2024wpp_corner.pdf}
% \caption{
% Corner plot of the posterior distributions for the radio light-curve model parameters of AT2024wpp. 
% The diagonal panels show the marginalized one-dimensional posteriors, and the off-diagonal panels show the joint posterior distributions. 
% Blue solid lines denote the median values, and black dashed lines indicate the 16th and 84th percentiles. 
% }
% \label{fig:corner_at2024wpp}
% \end{figure}

\end{document}